# Observation of Interaction-Induced Modulations of a Quantum Hall Liquid's Area

I. Sivan*, H. K. Choi*, A. Rosenblatt, Jinhong Park, Yuval Gefen, D. Mahalu, V. Umansky

**Studies of electronic interferometers, based on edge channel transport in the quantum Hall effect regime, have been stimulated by the search for evidence of abelian and non-abelian anyonic statistics of fractional charges. These studies found the ubiquitous electronic Fabry-Perot interferometer to be Coulomb dominated, thus masking coherent Aharonov-Bohm interference. Typically, the main signature of the Coulomb dominated regime is the lack of interference of the outer most edge channel as the magnetic field is varied. This seemingly surprising behavior is explained by the shrinkage of the interference area with increasing magnetic field, thus keeping the number of electrons and the enclosed flux constant. The model further stipulates – but with no experimental evidence thus far - that once the area shrinks by a size corresponding to an entire flux-quantum, it abruptly inflates to its original size leading to an unobservable $2\pi$ phase jump. Here we report on the observation of such area 'breathing' by performing a partial screening of the Coulomb interactions. The novelty is that the variation of the phase induced by the external knobs is not fully cancelled by the area response. We infer the area variations (with a resolution on the order of $\sim 50\ nm^2$) from conductance measurements. The latter are sensitive to the continuous phase variations, and reveal abrupt phase-jumps (smaller than $2\pi$). Based on our results, we propose a new experimental tool for probing anyonic statistics.**

## Introduction

The behavior of electrons and quasi-particles in mesoscopic systems stems from the combination of their wave-nature, particle-nature, and the effect of Coulomb interactions. Electronic Fabry-Perot interferometers (FPIs), and more generally ring-like geometries, have been utilized for the investigation of all these three facets[1–4] and the rich interplay among them[5–10]. In particular, electronic FPIs have possibly been the most studied candidates for probing Abelian and non-Abelian anyonic statistics of quasiparticles in the fractional quantum Hall effect (FQHE) regime[11–13]. Notwithstanding considerable efforts, the experimental study of anyonic statistics is still lagging behind theory, mostly owing to the adverse effect of Coulomb interactions. Two distinct regimes in the operation of FPIs - characterized by the conductance oscillation frequencies with the applied flux - have been reported[7,8] and studied theoretically[14–18]. In the coherent AB regime, one period of these oscillations corresponds to the increase of the threaded flux through the interferometer's area by a single flux quantum $\phi_0 = h/e$ ($e$ - electron charge, $h$ - Planck's constant)[19]. On the other hand, in the Coulomb-dominated (CD) regime, it is the mutual capacitance between the device and a modulation gate that dictates the oscillation periodicities. Interference in the AB regime is harder to observe, and has not been achieved thus far in any fractional quantum Hall state. At the same time, the CD regime has its own disadvantage - it does not and cannot reveal the anyonic statistics. Here, we report on a novel device which breaks this 'no go' state of affairs and potentially paves a new path for probing anyonic statistics.

The AB phase underlying interference patterns evolves as $2\pi\delta\phi_{AB}/\phi_0 = 2\pi(A\delta B + B\delta A_0)/\phi_0$, where $\delta\phi_{AB}$ is the variation of the threaded flux through the FPI, $B$ the applied magnetic field, $A$ the area enclosed by the interfering edge channel, and $\delta A_0 = \alpha \cdot \delta V_{MG}$ is the area variations due to the modulation gate (but excluding the effect of Coulomb interaction as explained below). Here $\alpha$ is the edge channel-modulation gate mutual capacitance. The AB oscillations with $\delta B$ and $\delta V_{MG}$ are hence characterized by the following frequencies:

$$1/\Delta B^{(AB)} = A/\phi_0 \quad , \quad (1a)$$

$$1/\Delta V_{MG}^{(AB)} = \alpha B/\phi_0 \ , \quad (1b)$$

By contrast, in the CD regime with the outer most edge channel interfering, there is no $B$ dependence of the conductance,

$$1/\Delta B^{(CD)} = 0 \ , \quad (2a)$$

$$1/\Delta V_{MG}^{(CD)} = \gamma \ . \quad (2b)$$

Here $1/\gamma$ is the voltage required to remove a single electron from the FPI[3-6], which does not depend upon the magnetic field (in similitude to the single electron transistor[20]).

Seemingly, these two regimes may seem to be of very different nature: while the AB oscillations are due to coherent interference, the CD oscillations reflect solely the electron occupancy of the device, and do not contain any information regarding the phase acquired by the interfering particles. This interpretation turned out to be too naïve. A unified theoretical framework has suggested that the CD oscillations may, in fact, be AB oscillations modified by Coulomb interaction[14]. According to this interpretation[7,14], as the magnetic field increases (decreases) so that the AB phase varies by $A\delta B$, the area will shrink (inflate) by $\delta A_{int}$ in order to keep a constant flux $A\delta B + B\delta A_{int} = 0$, where the subscript $int$ stands for 'interactions'. This variation of the area guarantees that the charge enclosed in the interfering area is kept constant, minimizing the charging energy of the device. Then, once the flux variation reaches the value of a whole flux quantum, $A\delta B = \phi_0$, an electron abruptly "jumps" from the bulk into the edge and the original area is restored[7,14].

According to this description, the area enclosed by the outer edge-channel is 'breathing'; upon increasing the magnetic field it continuously shrinks and then abruptly inflates, altogether with periodicity $\Delta B = \phi_0/A$. Nonetheless, the breathing in the CD regime has never been experimentally confirmed. The reason is that the measured conductance is simply constant as the magnetic field is varied. Basically, the area response to increasing the magnetic field fully compensates the effect of the latter. The subsequent abrupt addition of an electron leads to a $2\pi$ jump in the interference phase - clearly an unobservable effect as well. We are thus "blind" to this phenomenon of area breathing.

Here, we report on the realization of a novel device which combines the advantages of AB interferometers with CD response. It operates in an intermediate regime - in which the area response does not fully compensate the variations of the magnetic field, rendering the area "breathing" observable. In other words, $\delta A_{int} = -\xi \cdot \frac{A\delta B}{B}$, with $0 < \xi < 1$, leading to a total phase evolution $\delta\phi_{tot} = A\delta B + B\delta A_{int} = (1-\xi) \cdot A\delta B$. Measurements were performed on three different devices showing three distinct behaviors: AB-dominated, namely, $\xi = 0$ with no area breathing; Coulomb-dominated, namely, $\xi = 1$ with invisible breathing of the area; and an intermediate one ($\xi = 0.75$), with clear area breathing. We present conductance measurements of this novel device which incorporate both AB and CD frequencies (Eq. 1 & 2), hence contain information regarding both interference and interactions, and the interplay between them. Moreover, employing detailed analysis based on the system's energy, we deduce the device's charge stability diagram, providing an insight to its rich physics, being 'hidden' in the previously reported AB and CD devices. Furthermore, we provide proposals for utilizing our device for probing anyonic statistics of quasi-particles in the FQHE regime.

**Experimental setup**

The devices were fabricated on a GaAs-AlGaAs heterostructure, embedding a two-dimensional electron gas (2DEG) with electron density $\sim 2 \times 10^{11} \, cm^{-2}$. Electron-beam and optical lithography were employed in the fabrication process, and measurements were performed at electron temperature

~30 $mK$. All FPIs consist of a pair of quantum point contacts (QPCs), playing the role of semi-transparent mirrors (Fig. 1). Three different types of devices are shown in Fig 1. In Fig. 1a we show the simplest realization of a FPI; such FPIs have high charging energy regardless of their size[7,21], thus showing CD behavior. In order to suppress Coulomb interactions, a small (0.5 $\mu m^2$) ohmic contact is placed at the center of the interferometer's bulk[22] (Fig. 1b). Such devices, as well as top-gated ones[7,21,22], show AB behavior. A third type of device, unreported previously to the extent of our knowledge, consists on placing such ohmic-contacts in the vicinity of the FPI, as depicted in Fig. 1c. Unexpectedly, we found that in such devices, the suppression of the Coulomb interactions is made less effective, giving rise to a new regime intermediate between AB and CD. All three devices were fabricated on the same heterostructure and designed to have the same size $A \approx 2.5 \, \mu m^2$ so that they differ merely by the presence and the position of the ohmic-contacts. A fourth type of device, with the ohmic contact inside its bulk, and surrounded by an additional gate was found to operate in the intermediate regime as well, and is discussed in Supplementary Material (Supplementary Note 1 & Supplementary Figures 1-3). All measurements were conducted at bulk fillings $1 < \nu_B < 2$, with the outer most edge-channel interfering.

## Observations

Tuning the magnetic field to the $\nu_B = 2$ plateau and partitioning the outer edge channel, conductance oscillations are measured as function of $\delta B$ and $\delta V_{MG}$, with the different FPIs shown in Fig. 1. For the device shown in Fig. 1b, constant-phase lines in the $B$-$V_{MG}$ 2D plane follow a negative slope – a typical result in the AB regime (Fig. 2a). A fast Fourier transform (FFT) extracts $\phi_0/\Delta B^{(AB)} = 2.7 \, \mu m^2$ – agreeing well with the lithographically enclosed area, and $1/\Delta V_{MG}^{(AB)} = 103 \, mV^{-1}$ – needed to remove one flux quantum from the interior of the device (Eq. 1). For the device shown in Fig. 1a, no oscillation as function of $\delta B$ is observed – in agreement with the prediction for the CD regime (Eq. 2) - with a modulation gate frequency $1/\Delta V_{MG}^{(CD)} = 273 \, mV^{-1}$ – needed to remove one electron.

We then turned to employing the device in Fig. 1c. We commenced by measuring the conductance for rather pinched QPCs ($<t> = 0.05$), shown in Fig. 3a. This lattice-like conductance plot represents the charge stability diagram of the device and contains information both on the interference and on the device's charging, as we elaborate in Discussion. Although this plot is rather complex, its 2D FFT, shown in Fig. 3b, is simple; it reveals a lattice of peaks (as expected from an FFT of a lattice) all being linear combinations of underlying AB and CD frequencies, denoted on top of the figure. We find $\phi_0/\Delta B^{(AB)} = 2.6 \, \mu m^2$ & $1/\Delta V_{MG}^{(AB)} = 90 \, mV^{-1}$ for the AB frequencies, and

$1/\Delta B^{(CD)} = 0$ & $1/\Delta V_{MG}^{(CD)} = 222\ mV^{-1}$ for the CD frequencies. These frequencies are found relatively close to the ones of the 'pure AB' and 'pure CD' cases extracted from Fig. 2a & 2b, respectively. Furthermore, we find $1/\Delta V_{MG}^{(AB)}$ to be the only frequency that depends on magnetic field (in the Tesla range), as anticipated from Eq. 1 & 2 (see Supplementary Figure 4).

Moreover, by further opening the confining QPCs to $<t> = 0.7$ the presence of abrupt jumps in the conductance (*phase jumps*) becomes more clear (Fig. 4a), which shows good agreement with the conductance's model which stems from the minimization of the system's energy (Fig. 4b, see Discussion below and Supplementary Notes 2-3).

# Discussion

Since the AB and CD types of behavior are merely two different regimes of operation of the same device, they may be analyzed within the same theoretical framework[14]. We address a setup where transport through the outer most edge channel at $1 < \nu_B < 2$ gives rise to interference. This case has been studied experimentally. Hereafter, we refer to the interfering outer most edge channel as the "edge"; and to the trapped puddle of charge at the center of the FPI (the partially filled second Landau level) as "bulk"[23]. The charge added to the edge (and thereby to the lowest Landau level) is denoted $\delta Q_{edge}$ and the charge added to the bulk is denoted $\delta Q_{bulk}$.

We consider a minimal capacitive model[1,3,4] for describing the FPI setup, depicted in Fig. 1d. The model consists of three capacitances $C_{edge}, C_{bulk}, C_{eb}$ describing the mutual Coulomb interactions between the system's three components $\delta Q_{edge}, \delta Q_{bulk}$ & $\delta V_{MG}$. The total energy of the system can then be written as the sum of three contributions[14,16,24,25], $E_{total} = E_{eb} + E_{edge} + E_{bulk}$, where the second and third terms are the edge and bulk charging energies (depending solely on $\delta Q_{edge}$ and $\delta Q_{bulk}$ respectively), and the first term is due to the interaction between them. This first term can be presented as function of their sum $\delta Q_{edge} + \delta Q_{bulk}$. The relation between this approach and an earlier one (cf. Ref. 5), where the variable $\delta Q_{edge} \cdot \delta Q_{bulk}$ have been employed, is discussed in Supplementary Note 2. The system's charge state $(\delta Q_{edge}, \delta Q_{bulk})$ is determined by minimizing the total energy $E_{total}$.

In the quantum Hall regime, the charge, $\delta Q_{edge}$, added to the edge follows the total flux, $\delta \phi_{tot}$, enclosed by the latter: as the flux increases (decreases), single-particle states cross in below (cross in above) the Fermi energy, increasing (decreasing) the number of occupied states on the edge:

$$\delta Q_{edge} = e \cdot \frac{\delta \phi_{tot}}{\phi_0} = e \cdot \frac{A\delta B + B\delta A_0 + B\delta A_{int}}{\phi_0} \ . \quad (3)$$

Here, the third term on the r.h.s represents the interaction-induced area-response. Thus the overall variation of the area $\delta A = \delta A_0 + \delta A_{int}$ is separated into two components. The first component is a linear component which is solely a result of variations of the modulation-gate voltage, $\delta A_0 = \alpha B \delta V_{MG}$, and stems solely from its mutual capacitance to the interfering edge. The second component is an oscillating component[7,14], $\delta A_{int}$, which is a function of both $\delta B$ and $\delta V_{MG}$, for which we derive the full expression below.

**Magnetic field variations**

As the magnetic field increases, and the filling factor decreases, charge is removed from the higher Landau level (and thereby from the "bulk") into the lowest Landau level (and thereby to the "edge"). More precisely, with every flux quantum added, the degeneracy each Landau level increases by one; resulting in a transfer of a single electron from the second Landau level (thus $\delta Q_{bulk} = \pm 1$) to the lowest Landau level (thus $\delta Q_{edge} = \mp 1$), while keeping the total charge in the system constant. The corresponding variation of energy[14,15]:

$$\delta E_{total}(\delta B) = \frac{K_{EB}}{2} \cdot (\delta Q_{tot})^2 + \frac{K_E}{2} \cdot \left(\delta Q_{edge} - e \cdot \frac{A\delta B}{\phi_0}\right)^2 + \frac{K_B}{2} \cdot \left(\delta Q_{bulk} + e \cdot \frac{A\delta B}{\phi_0}\right)^2, \quad (4)$$

where $\delta Q_{tot} = \delta Q_{edge} + \delta Q_{bulk}$, and $K_{EB}, K_E, K_B$ (associated with effective capacitances) are system specific. In particular, $K_{EB}$ depends on the Coulomb coupling between the edge and the bulk[14].

Eq. 4 implies that the energy is minimized if $\delta Q_{edge} = e \cdot \frac{A\delta B}{\phi_0}$ and $\delta Q_{bulk} = -e \cdot \frac{A\delta B}{\phi_0}$, leading to $\delta Q_{tot} = 0$ at all times. This continuous charging of the edge, $\delta Q_{edge}$, should result in the periodicity $\Delta B = \phi_0/A$. The latter coincides with the behavior in the AB regime, as seen in Fig. 2a, measured with the device shown in Fig. 1b. In this device, while $\delta Q_{edge}$ is continuously charged from the leads, $\delta Q_{bulk}$ is discharged into the center ohmic-contact.

Previous measurements have also reported the observations of AB oscillations even without the center ohmic-contact, but rather with a top-gate covering the device[7,21]. In that case, the bulk charge $\delta Q_{bulk}$ cannot vary continuously (as in the case with an ohmic contact at the center of the device) but rather discretely. This is because the bulk is isolated by an incompressible strip[14,23]. Thus, as the magnetic field increases such that $A\delta B < \phi_0$, the edge is continuously charged by $\delta Q_{edge}^{(AB)}(\delta B) = e \cdot$

$\frac{A\delta B}{\phi_0}$, leading in this case to an increase of the total charge $\delta Q_{tot}^{(AB)}(\delta B) = \delta Q_{edge}^{(AB)}(\delta B) = e \cdot \frac{A\delta B}{\phi_0}$ with $\delta Q_{bulk} = 0$ due to the bulk's charge quantization. This of course will increase the charging energy (which favors $\delta Q_{tot} = 0$; first term in Eq. 4), which is possible since the top-gate effectively decreases $K_{EB}$ such that $K_{EB} \ll K_E$ [7,21,14]. Then, once a whole flux quantum is added $A\delta B = \phi_0$, the energy can again be minimized by discharging the bulk abruptly $\delta Q_{bulk} = -e$. This AB behavior is summarized in Figs. 5a-5c (blue), leading to a conductance depicted in Fig. 5d (blue).

On the other hand, as alluded above in the CD regime, the total charge must be constant at all times. This is due to the large charging energy $K_{EB} \gg K_E$ (in the absence of an ohmic-contact or top-gate). As a result, when the applied magnetic field is increased, the area shrinks in order to prevent the enclosed flux from increasing. This means $\delta Q_{tot}^{(CD)}(\delta B) = \delta Q_{edge}^{(CD)}(\delta B) = e \cdot (A\delta B + B\delta A_{int})/\phi_0 = 0$, with $\delta Q_{bulk} = 0$ due to the bulk's quantization, leading to $\delta A_{int} = -A\delta B/B$. In that case, as the interferometer's area shrinks, a dipole is created: a depleted region (positively charged) is created just outside the boundary of the shrinking interferometer, while an excess negative charge accumulates inside. The energy increase due to this dipole is reflected by the second and third terms in Eq. 4. Once a whole flux quantum is added $A\delta B = \phi_0$, the energy can again be minimized by abruptly moving an electron from the bulk to the edge. This is described by the simultaneous variation $\delta Q_{bulk} = -e$ and $\delta Q_{edge} = +e$. At this point, the area response $\delta A_{int}$ should abruptly vanish from ~810 $nm^2$ (a flux quantum's area at 5 $T$) to 0 $\mu m^2$. This abrupt response cannot be inferred from the measured conductance (Fig. 2b) since it should result in a $2\pi$ variation of the Aharonov-Bohm phase. This leaves the conductance constant as function of $\delta B$. This CD behavior is summarized in Figs. 5a-5c (purple), leading to a conductance depicted in Fig. 5d (purple).

To summarize, for any device without an ohmic-contact within its center (implying a quantization condition on $\delta Q_{bulk}$) the system "decides" whether it shrinks or not, and by how much, according to the values of $K_{EB}$ and $K_E$. Taking the derivative of the total energy with respect to $\delta A_{int}$ we obtain $\delta A_{int} = -\xi \cdot A\delta B/B$, with $\xi = \frac{K_{EB}}{K_E + K_{EB}}$. The AB regime corresponds to the limit $K_E \gg K_{EB}$, leading to $\xi = 0$ and $\delta A_{int} = 0$; the CD corresponds to the limit $K_E \ll K_{EB}$, leading to $\xi = 1$ and $\delta A_{int} = -A\delta B/B$. This behavior in the intermediate regime is summarized in Figs. 5a-5c (red and yellow for $\xi = 0.25$ and $\xi = 0.75$ respectively), resulting in the conductance depicted in Fig. 5d.

The theoretical prediction for the conductance ($\xi = 0.75$; Fig. 5d, yellow) is consistent with the measured signal shown in Fig. 4c, from which we have retrieved both the area = 2.6 $\mu m^2$ (from the main periodicity), and the interaction parameter $\xi = 0.75$ (from the amplitude of the phase jumps).

The possibility to extract these two parameters from a single magnetic field scan demonstrates the versatile information contained in the intermediate regime - it incorporates both interactions (quantified by $\xi$) and interference (related to $A$). Furthermore, from the conductance (Fig. 4c) we infer the evolution of the area response, $\delta A_{int}$, as function of $\delta B$, shown in Fig. 4d. We find a maximal area response of $\delta A_{int} = 500\ nm^2$, which is in good agreement with our expectation $\delta A_{int} = -\xi \cdot \frac{\phi_0}{B} = 0.75 \cdot 585\ nm^2 \approx 440\ nm^2$.

**Modulation gate variations**

The two handles we have, varying the magnetic field $\delta B$ and varying the modulation gate voltage $\delta V_{MG}$, are substantially different from each other. While in both AB and CD regimes the total charge must be conserved for $\delta B$ variations on the scale of several flux quanta, $A\delta B \gg \phi_0$, this is not the case for $\delta V_{MG}$, which induces charge on the device through its capacitance $\gamma$: applying $\delta V_{MG}$ to the modulation-gate requires charging the device by $\delta Q_{tot} = \gamma \delta V_{MG}$ which divides into[14]: $\delta Q_{edge} = e \cdot \frac{B\delta A_0}{\phi_0}$ and $\delta Q_{bulk} = \gamma \delta V_{MG} - e \cdot \frac{B\delta A_0}{\phi_0}$. It follows that the energy of the system (Eq. 4) can be reformulated as[14]:

$$\delta E_{total} = \frac{K_{EB}}{2} \cdot (\delta Q_{tot} - \gamma \delta V_{MG})^2 + \frac{K_E}{2} \cdot \left(\delta Q_{edge} - e \cdot \frac{\delta \phi_{AB}}{\phi_0}\right)^2$$

$$+ \frac{K_B}{2} \cdot \left(\delta Q_{bulk} - \left(\gamma \delta V_{MG} - e \cdot \frac{\delta \phi_{AB}}{\phi_0}\right)\right)^2. \quad (5)$$

Although Eq. 5 is valid for any variation of $\delta B$ and $\delta V_{MG}$, only modulation-gate variations, $\delta V_{MG}$, are considered in this section.

In the AB regime, as the modulation-gate voltage increases by $\delta V_{MG}$ such that $\delta \phi_{AB}(\delta V_{MG}) = B\delta A_0 = \alpha B \delta V_{MG}$, the edge is continuously charged by $\delta Q_{edge}^{(AB)}(\delta V_{MG}) = e \cdot \frac{\alpha B \delta V_{MG}}{\phi_0}$ allowing us to observe Aharonov-Bohm interference with a period $\Delta V_{MG}^{(AB)} = \frac{\phi_0}{\alpha B}$ (cf. Fig. 2a). Note that $\alpha$ is independent to the magnetic field (Supplementary Figure 4). Simultaneously, we expect the bulk to be continuously charged from the center ohmic-contact, satisfying $\delta Q_{bulk}^{(AB)}(\delta V_{MG}) = \gamma \delta V_{MG} - e \cdot \frac{\alpha B \delta V_{MG}}{\phi_0}$. Alternatively, with top-gated devices ($K_{EB} \ll K_E$) and quantization of the bulk charge, the bulk acquires an additional electron discretely only once this last expression (which can be regarded as the induced charge in the bulk) amounts to an increase by a whole electron charge $\gamma \delta V_{MG} - e \cdot$

$\frac{\alpha B \delta V_{MG}}{\phi_0} = e$. This AB behavior is summarized in Figs. 5e-5g (blue) and results in a conductance depicted in Fig. 5h (blue).

By contrast, in the CD regime, the high charging energy $K_{EB} \ll K_E$, requires $\delta Q_{tot}^{(CD)}(\delta V_{MG}) = \gamma \delta V_{MG}$ for all values of $\delta V_{MG}$. As $\delta Q_{bulk} = 0$ due to the bulk charge quantization, it follows that $\delta Q_{edge}^{(CD)}(\delta V_{MG}) = \gamma \delta V_{MG}$. According to Eq. 3 this necessitates an area response $\frac{B \delta A_{int}}{\phi_0} = \frac{\gamma \delta V_{MG}}{e} - \frac{\alpha B \delta V_{MG}}{\phi_0}$, (which is always larger than zero since $\frac{\gamma}{e} > \frac{\alpha B}{\phi_0}$, see Supplementary Figures 2 & 4). Namely, while in the AB regime the area dilations are dictated solely by the mutual capacitance between the edge and the modulation-gate, $\alpha$, here, in the CD regime, the area dilations overshoots in order to satisfy the device's capacitance to the modulation-gate, $\gamma$. This overshoot will cost an energy proportional to $K_E$ (Eq. 5). Then again, once this overshoot is equivalent to a whole electron, the area will retract back minimizing the total energy. This CD behavior is summarized in Figs. 5e-5g (purple) and results in a conductance depicted in Fig. 5h (purple).

Similarly to the analysis of magnetic field variations, we can express the area response to variations of $\delta V_{MG}$ for any value of the interaction parameter $\xi$: we obtain $\frac{B \delta A_{int}}{\phi_0} = \xi \left( \frac{\gamma \delta V_{MG}}{e} - \frac{\alpha B \delta V_{MG}}{\phi_0} \right)$ resulting in a total area variation $\frac{B \delta A}{\phi_0} = \frac{\alpha B \delta V_{MG}}{\phi_0} + \xi \left( \frac{\gamma \delta V_{MG}}{e} - \frac{\alpha B \delta V_{MG}}{\phi_0} \right)$. This is summarized in Figs. 5e-5h (red and yellow for $\xi = 0.25$ and $\xi = 0.75$ respectively).

## Analysis of the entire $B - V_{MG}$ plane

So far we have analyzed the area response to variations of either $\delta B$ or $\delta V_{MG}$. We now turn our attention to concomitant variations of both. This will allow us to analyze the data shown in Figs. 3a & 4a. Notably, the interference pattern in Fig. 3a & 4a may be decomposed into two ingredients. First, descending lines along which phase-jumps take place (cf. solid lines in Fig. 4a). The respective periodicities are denoted $\Delta V_{MG}^{(PJ)}$ and $\Delta B^{(PJ)}$ (marked on top of Fig. 4a, with the superscript $PJ$ for 'phase jump'). Second, continuous conductance oscillations with respect to modulation gate voltage and magnetic field in between two adjacent phase jump lines. The corresponding periodicities are then $\Delta V_{MG}^{(m)}$ and $\Delta B^{(m)}$ respectively (marked on top of Fig. 4a as well, with the superscript $m$ for 'modified'). These continuous oscillations reflect the interference of electrons at the edge. Crossing a maximal conductance line (e.g. following arrow $a_2$ in Fig. 3a) corresponds to an incremental variation of $\delta Q_{edge}$ by $\pm e$. Here, 'modified' alludes to the fact that these continuous oscillations are in fact coherent AB oscillations modified by the Coulomb interaction. Likewise, crossing a phase-jump line (e.g. following arrow $a_1$ in Fig. 3a) implies a change of the bulk charge $\delta Q_{bulk}$ by $\pm e$. Naturally, the combination of these two types of processes gives rise to the system's charge stability diagram in the $B - V_{MG}$ plane.

In order to express the modified periodicities $\Delta B^{(m)}$ & $\Delta V_{MG}^{(m)}$ (in between abrupt changes of the charge) we need to determine an expression for the interference phase (Eq. 3) for a fixed number of electrons in the bulk. We first generalize the dependence of the area response on $\delta B$ and $\delta V_{MG}$ by combining the results obtained in the previous sections: $\frac{B\delta A_{int}}{\phi_0} = \xi \left( \frac{\gamma \delta V_{MG}}{e} - \frac{\delta \phi_{AB}}{\phi_0} \right) + \xi \cdot \frac{\delta Q_{bulk}}{e}$, where the last term represents phase jumps which accompany variations of $\delta Q_{bulk} = \pm e$ (for further details see Supplementary Notes 2-3). This result is then plugged into the expression for the total phase (Eq. 3)[14]:

$$2\pi \frac{\delta \phi_{tot}}{\phi_0} = 2\pi \left[ (1-\xi) \cdot \frac{\delta \phi_{AB}}{\phi_0} + \xi \cdot \frac{\gamma \delta V_{MG}}{e} + \xi \cdot \frac{\delta Q_{bulk}}{e} \right]. \quad (6)$$

Note that the last term of the of the equation accounts for abrupt phase jumps. The respective modified periodicities are:

$$1/\Delta B^{(m)} = (1-\xi) \cdot 1/\Delta B^{(AB)} \quad . \quad (7a)$$

$$1/\Delta V_{MG}^{(m)} = (1-\xi) \cdot 1/\Delta V_{MG}^{(AB)} + \xi \cdot 1/\Delta V_{MG}^{(CD)} \quad . \quad (7b)$$

These two frequencies are notably a linear combination of the AB and CD frequencies (cf. Eqs. 1 & 2): $1/\Delta V_{MG}^{(AB)} < 1/\Delta V_{MG}^{(m)} < 1/\Delta V_{MG}^{(CD)}$ and $1/\Delta B^{(CD)} < 1/\Delta B^{(m)} < 1/\Delta B^{(AB)}$. Constant flux lines of $\delta \phi_{tot}$ follow a negative slope in the $B - V_{MG}$ plane, similar to the pure-AB case. We extract $\Delta B^{(m)} = 65.9\ G$ & $\Delta V_{MG}^{(m)} = 5.3\ mV$ from our data; each of these periodicities leads to an interaction parameter $\xi = 0.75 \pm 0.01$. The periodicities associated with phase jumps are found to be $1/\Delta B^{(PJ)} = A/\phi_0$ and $1/\Delta V_{PJ}^{(PJ)} = \gamma - \alpha B/\phi_0$ (cf. Eqs. 1 & 2; see Supplementary Note 3 for details). We find this relation to be in agreement with the measured values. Note that these periodicities do not depend on the value of $\xi$.

Combining the modified periodicities (Eq. 7) with the phase jump periodicities allows us to construct the charge stability diagram. Naturally, vectors connecting different cells in the diagram (Fig. 3a) represent different discrete charge variations $(\delta Q_{edge}, \delta Q_{bulk}) = (n, m)$ (cf. Supplementary Note 2). It is convenient to select the following basis (Fig. 3a): $\boldsymbol{a}_1 = (\Delta B^{(AB)}, 0)$ and $\boldsymbol{a}_2 = \left( \Delta B^{(AB)} \left( 1 - \right. \right.$

$\frac{\Delta V_{CD}}{\Delta V_{AB}}$), $\Delta V_{MG}^{(CD)}$). Here, $\boldsymbol{a_1}$ represents the process of moving one electron from the bulk to the edge (namely, $n = 1, m = -1$, as explained in 'magnetic field variations'), while $\boldsymbol{a_2}$ represents adding one electron to the edge keeping the bulk charge constant (namely, $n = 1, m = 0$). Evidently, their sum $\boldsymbol{a_1} + \boldsymbol{a_2}$ refers to adding one electron to the bulk while keeping the edge charge constant. The reciprocal lattice is spanned by the pure AB and CD frequencies, $\boldsymbol{b_1} = 2\pi \cdot (1/\Delta B^{(AB)}, 1/\Delta V_{MG}^{(AB)})$ and $\boldsymbol{b_2} = 2\pi \cdot (0, 1/\Delta V_{MG}^{(CD)})$ respectively (cf. Fig. 3b).

**Phase jumps and proposed experiments**

Finally, we study qualitatively the phase jump taking place when a phase-jump line is crossed, following an infinitesimal variation of either $\delta B$ or $\delta V_{MG}$. The acquired phase is given by[14]:

$$\Delta \theta = -2\pi \cdot \xi \ , \quad (8)$$

Evidently, in the extreme AB and CD cases this phase jump is zero ($\xi = 0$) or unobservable ($\xi = 1$) respectively. Experimentally, the value of $\xi$ ($0 < \xi < 1$) can be deduced from directly measuring the phase jump (seen in Fig. 4c). We find $\xi \approx 0.75$ (similar to the value we obtained independently by comparing the measured frequencies with Eq. 7 of the previous section). This value of $\xi$ corresponds to an area jump of $440 \ nm^2$, cf. Fig. 4d.

A possible extension of our analysis to the FQHE regime is straightforward and results in a phase jump given by:

$$\Delta \theta = \theta_{stat} \cdot (1 - \xi) \ , \quad (9)$$

where $\theta_{stat}$ is the quasi-particles' statistical braiding phase (e.g $\theta_{stat} = \frac{2\pi}{3}$ for anyons at $\nu_B = 1/3$ ). While the difference between Eq. 8 and Eq. 9 is minute, its impact might be significant. Previously, in the search for anyonic statistics, pure AB ($\xi = 0$) was sought in order to reveal $\theta_{stat}$ (only pure AB and pure CD were experimentally available). Now, utilizing our new devices, the requirement for probing $\theta_{stat}$ is substantially less restrictive; in fact, any value $\xi < 1$ should suffice.

# Summary

We provide here an experimental evidence of area modulations of a FPI implemented in the quantum Hall effect regime. These modulations stem from the minimization of the system's energy as the applied magnetic field and the modulation gate voltage are varied. Our analysis does not rely on the coherent operation of the device. The measurement of the area modulations, though, could be realized through the observation of coherent interference oscillations. The area evolution consists of a continuous shrinking when increasing magnetic field, followed by an abrupt dilation.

We have employed a theoretical framework which accounts for both the AB, the CD, and a novel intermediate regime. The latter enables us to construct the system's charge stability diagram as well showing good agreement with a minimal theoretical model. Both the area breathing and the charge stability diagram are used towards a complete characterization of the device, the latter being hidden in the pure AB or CD limits.

Utilizing a similar interferometer operation in the intermediate regime in the domain of the fractional quantum Hall effect regime is called for. Previous attempts to employ FPIs in the FQHE regime have utilized both large area and small area devices. For the former, Coulomb effects are expected to be negligible if a top-gate is placed on top of the sample or an ohmic contact is placed at the center of the device. However, AB interference (observable in the integer quantum Hall effect regime) has not been observed for fractional filling factors. This may be attributed to a short coherence length of anyons. One may then resort to smaller FPIs. Unfortunately, reducing the device's size leads to the undesired CD regime in the case of top-gated devices, and, in the case of devices with ohmic contacts at the center, further miniaturization is limited. The device reported here, having an ohmic contact placed outside its circumference, operates in a regime intermediate between AB and CD. Given the possibility to make the device's size smaller, yet maintaining $\xi < 1$ with a clear signature of coherent Aharonov-Bohm oscillations, presents us with an intriguing possibility: notwithstanding that the quasi-particle coherence length may be short, coherency is still maintained over the size of the interferometer. Interference signal, inductive of anyonic statistics, is then potentially observable. In other words, our findings reported here pave a new path for probing anyonic statistics in the FQHE regime.

# Methods

## Ohmic contacts

As explained in the text, the 'pure' CD regime can be avoided through undermining the effect of Coulomb interactions. This can be achieved utilizing two different methods:

First: We place a grounded ohmic contact inside the interferometer's bulk, thus avoiding its charge quantization (cf. the device in Fig. 1b). The ohmic contact consists of 106nm Au / 53nm Ge / 40nm Ni alloyed to the heterostructure. This OC is resistively coupled solely to the bulk (order of magnitude for the resistance to the inner channel at $\nu_B = 2$ is a few hundreds ohms), and clearly not to the edge (the observed coherent phase oscillations are evidence to that). This has also been verified in previous experiments by measuring the current flowing into the ohmic-contact[22].

Second: This consists of lowering the device's charging energy, $K_{EB}$. For this purpose, two schemes are available. One scheme consists of covering the whole area of the FPI by a metallic top gate, increasing its capacitance, thus lowering its charging energy[7,21]. This scheme has not been utilized in our measurements, but has been reported to work well for relatively large FPIs (areas larger than 12μm$^2$), resulting in a 'pure' AB behavior. On the other hand, devices of area smaller than 4μm$^2$ have been found to operate in the CD regime, even if covered by top-gate[7,21]. This was attributed to the different scaling of the capacitances with diameter[14]. The other scheme, utilized here for the first time, relies on placing an OC in the vicinity of the interferometer. Here we employ it not for its resistive coupling but for the capacitance it induces, partially screening the Coulomb interaction. As is discussed in the main text, this gives rise to the intermediate regime.

## Measurement techniques

Our electronic setup for conductance measurements is depicted in the Fig. 1c. We note that two amplifiers were used in the course of our measurements; the first, a home-made voltage preamplifier at *T*=1K having a gain factor ∼10 and a commercial amplifier (NF SA-220F5) at room temperature having a voltage gain ∼200.

All experiments are performed within $^3$He-$^4$He dilution fridges. Electron temperatures are measured via shot-noise measurements by driving a variable DC source current, at a center frequency of 800kHz and bandwidth of 10kHz

**Acknowledgements** We thank Moty Heiblum for immeasurable support and consulting with regards to all aspects of the undertaken work. We also thank Nissim Ofek for very useful comments and ideas; and Yonatan Cohen for sharing his insights and advice. We acknowledge the partial support of DFG Grant No. RO 2247/8-1 as well as the Minerva foundation, the U.S.-Israel Bi-National Science Foundation (BSF), the Israeli-German Foundation (GIF) and the European Research Council under the European Community's Seventh Framework Program (FP7/2007-2013)/ERC Grant agreement No. 227716.

**Author Contributions** I. S., H. K. C., J. P. and Y. G. contributed to paper writing. I. S., H. K. C. and A. R. contributed to sample design, device fabrication and data acquisition. I. S., H. K. C., J. P. and Y. G. contributed to the theoretical analysis. D. M. contributed to electron beam lithography. V. U. grew the heterostructures.

**Author information**. Correspondence and requests for materials should be addressed to I.S. (itamar.sivan@weizmann.ac.il) or H. K. C (hkchoi@weizmann.ac.il).

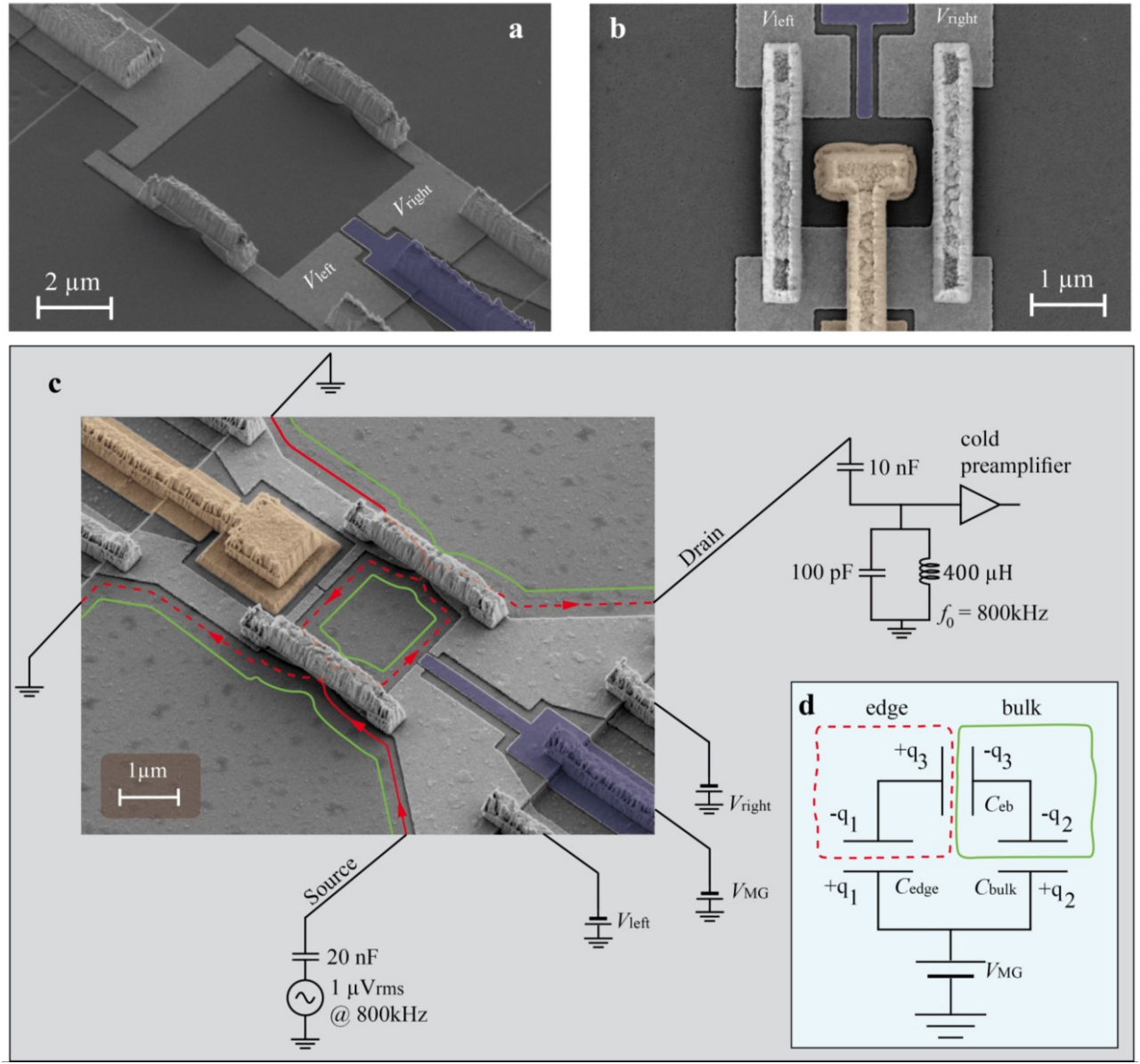

**Figure 1: SEM images and an illustration of the experimental setup. a**, Bare FPI; such devices show distinct Coulomb dominated (CD) behavior. **b**, FPI with a grounded ohmic contact in its center (gold, false-color) aimed at screening Coulomb interactions; such devices show Aharonov-Bohm (AB) interference. **c**, FPI with a grounded ohmic contact in its close proximity (gold, false-color) inducing partial screening of Coulomb interactions; such devices show intermediate behavior between the CD and AB. Red lines represent edge states and arrows represent the current's chirality. Current is injected into the device from the source, partitioned at the two QPCs, and probed at the drain employing a cold-amplifier. Partitioned current is denoted by dashed lines. Modulation gates are marked by blue (false-color) in all images.

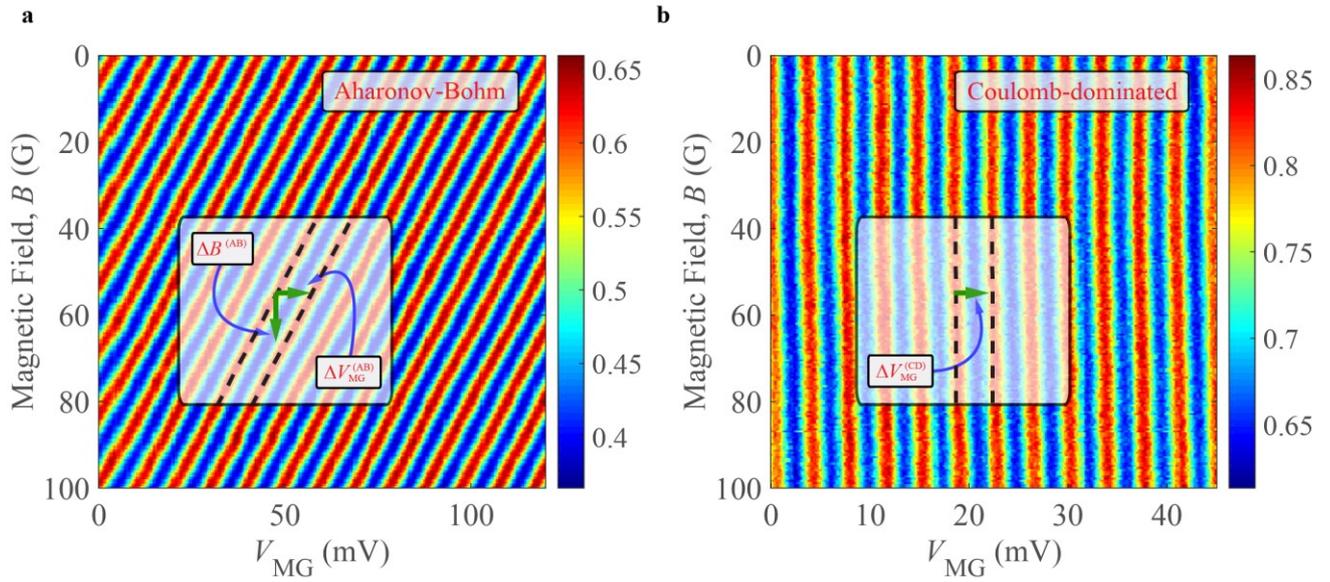

**Figure 2: Conductance as function of the modulation-gate voltage $\delta V_{MG}$ and the magnetic field $\delta B$, measured with Aharonov-Bohm dominated and Coulomb-dominated devices. a**, 2D conductance measured with the FPI with an ohmic-contact in its center (seen in Fig. 1b) at $B = 4.7T$. This plot shows a clear AB behavior: constant-phase lines are marked with grey lines and the AB periods $\Delta V_{MG}^{(AB)}$ & $\Delta B^{(AB)}$ are marked accordingly. **b**, 2D conductance measured with the bare FPI (seen in Fig. 1a) ) at $B = 4.7T$. This plot shows a clear CD behavior with no dependence of the conductance on magnetic field; constant-phase lines are marked with grey lines and the CD period $\Delta V_{MG}^{(CD)}$ is marked accordingly.

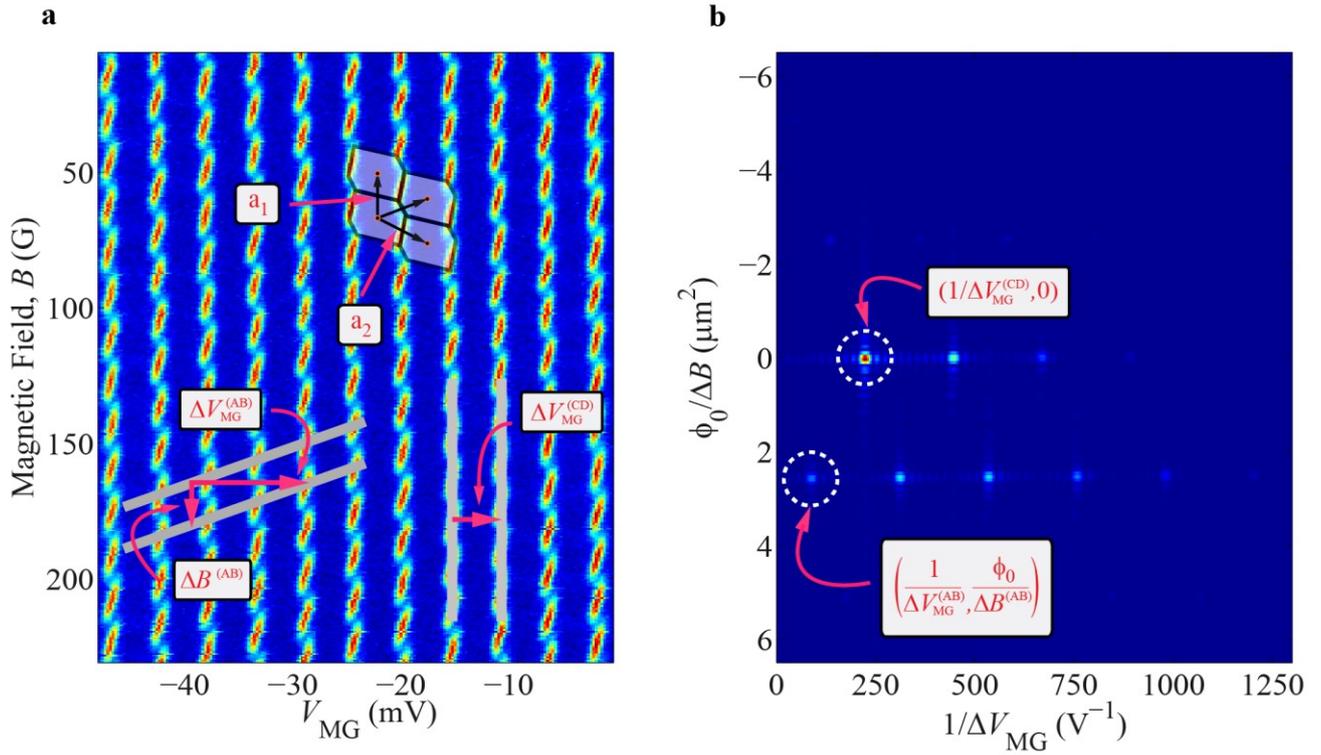

**Figure 3: Conductance of a FPI operating in an intermediate regime, as function of the modulation-gate voltage $\delta V_{MG}$ and the magnetic field $\delta B$. a**, 2D conductance measured with the FPI with an ohmic-contact in its vicinity (seen in Fig. 1c) at $B = 5T$, giving rise to a charge stability diagram. The vectors marked on top of the figure connect different cell and can be translated to physical processes, as elaborated in Discussion. Grey lines mark the underlying AB and CD frequencies. **b**, 2D FFT of the conductance in **a**, revealing a clear lattice structure as well, as expected from an FFT of a lattice. Underlying AB and CD frequencies are denoted by circles, and the reason for their appearance here is elaborated in Discussion.

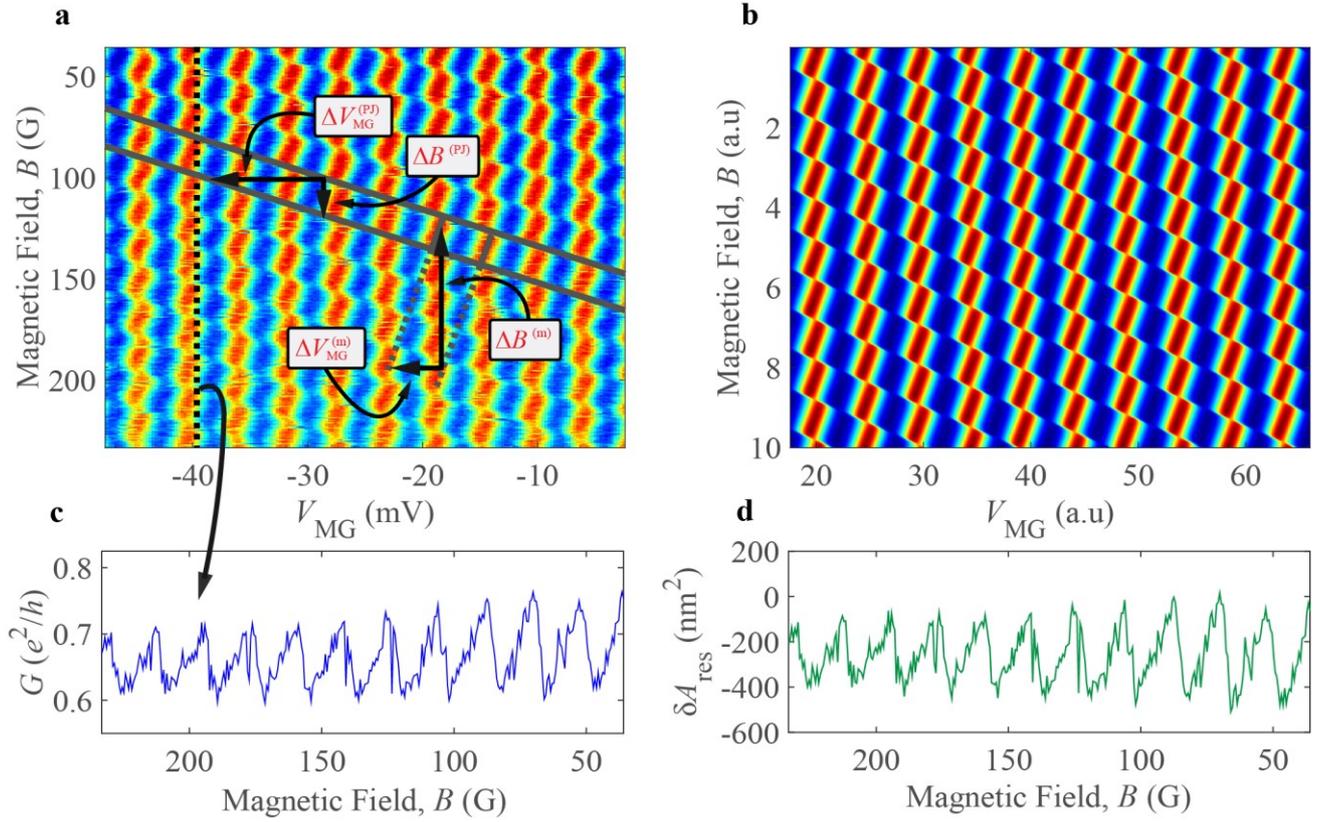

**Figure 4**: **2D conductance plots with clear phase-jump lines. a**, Conductance is measured with the same device as Fig. 3, but with substantially higher transmission ($<t> \approx 0.7$). Phase-jump lines are marked by solid descending lines and the frequency defined by their slope is marked by $\Delta V_{MG}^{(PJ)}$ & $\Delta B^{(PJ)}$. In between adjacent phase jump lines we have continuous conductance oscillations. By extrapolating these continuous oscillations (dashed ascending lines) we may attribute them periodicities denoted $\Delta V_{MG}^{(m)}$ & $\Delta B^{(m)}$. More details are provided in Discussion. **b**, Theoretical plot, simulated according to the conductance $G$ expressed in Discussion. **c**, Constant-$\delta V_{MG}$ line taken from **a**, showing clear phase accumulation followed by abrupt phase-jump of $\frac{\Delta \theta}{2\pi} = 0.75$, which is also equal to the value of $\xi$ (Eq. 8). **d**, The area response $\delta A_{int}$ inferred from the conductance' phase **c**. The relation between the area response and the phase is obtained by inverting the relation between $\delta A_{int}$ and the conductance $G$, provided in the caption of Fig. 5. Note that indeed Fig. 4c & 4d are highly similar (as anticipated by the theoretical plots, cf. yellow curves in Fig. 5d & 5b) but they are not identical. Naturally, for small values of the phase, the conductance is nearly proportional to it.

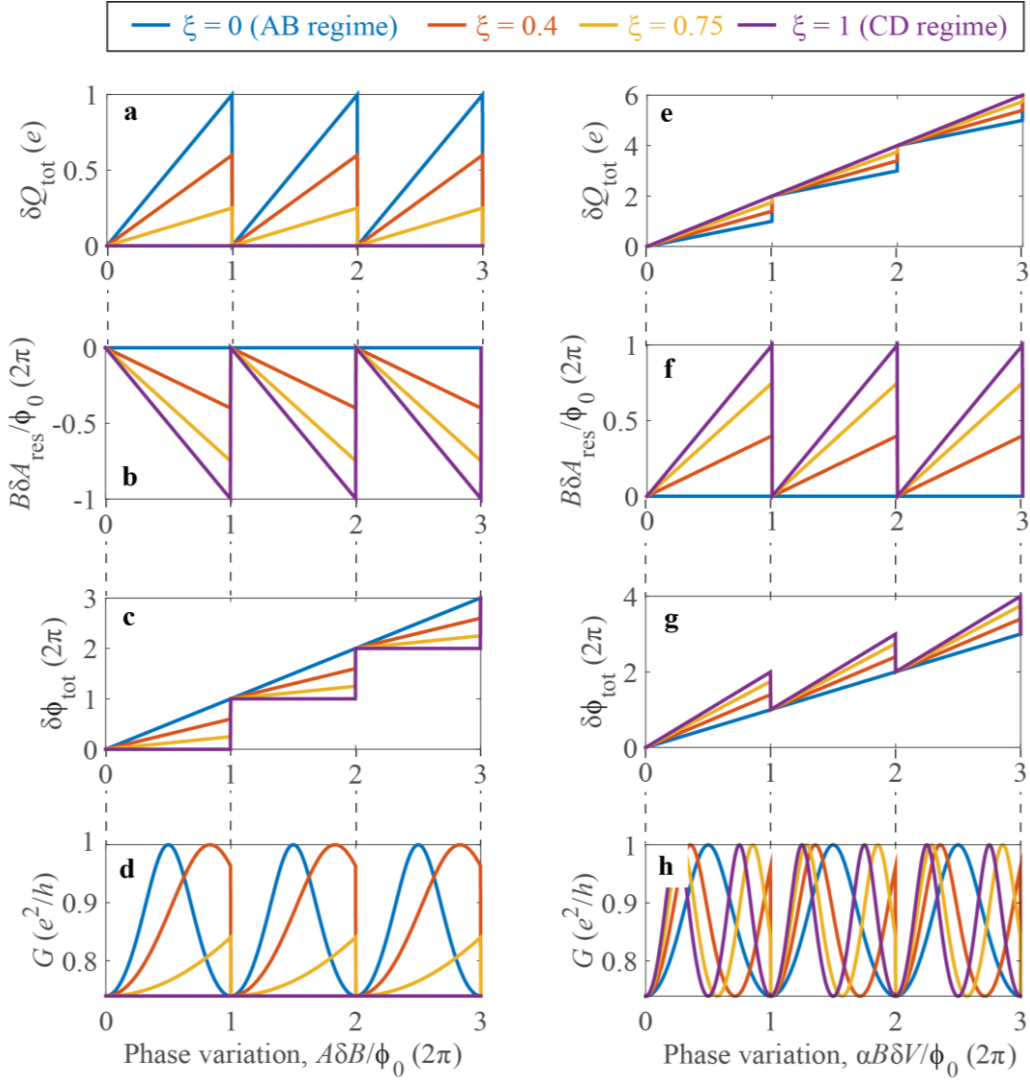

**Figure 5: Theoretical plots for the Fabry-Perot interferometer in all regimes. a**, Total charge in the system, $\delta Q_{tot} = \delta Q_{edge} + \delta Q_{bulk}$. **b**, The area response $\delta A_{int}$. **c**, The total flux $\delta \phi_{tot}$, including both the externally varied phase $A\delta B/\phi_0$ and the response $B\delta A_{int}/\phi_0$. **d**, The resulting conductance for relatively open QPCs ($t^2 = 0.85$), taking the straightforward form $G = f_0 + f_1 \cos\left(2\pi \cdot \frac{A\delta B + B\delta A_{int}}{\phi_0}\right)$ (further details are provided in Supplementary Note 3). All four graphs (**a-d**) are plotted versus variations of the magnetic field. **e, f, g, h**, Similar graphs as a, b, c & d, as a function of modulation gate voltage $\delta V_{MG}$. The plots are given for the simple case of $\nu_B = 2$.

# Observation of Interaction-Induced Modulations of a Quantum Hall Liquid's Area - Supplementary Materials

I. Sivan[*], H. K. Choi[*], A. Rosenblatt, Jinhong Park, Yuval Gefen, D. Mahalu, V. Umansky

**Supplementary Notes**

**Supplementary Note 1. FPI with a center ohmic-contact surrounded by gate**

An additional type of interferometer was found to show an intermediate regime between pure Aharonov-Bohm (AB) and Coulomb Dominated (CD). This interferometer, shown in Supplementary Figure 1 has an ohmic contact inside its bulk, which is surrounded by an additional gate. Supplementary Figures 2a,c show conductance measurements with respect to magnetic field and modulation gate (MG). The first conductance measurement was performed while the isolation-gate around the center ohmic-contact was unbiased; giving rise to simple AB oscillations (pajama-like stripe) following Eq. 1 (Supplementary Figures 2a). Its FFT reveals a single 2D-frequency with $\phi_0/\Delta B^{(AB)} = 12.7 \, \mu m^2$ and $\Delta V_{MG}^{(AB)} = 4.9 \, mV$ (Supplementary Figure 2b). The second (Supplementary Figures 2c) was performed while the isolation-gate was negatively biased so that the 2DEG beneath it is depleted (Supplementary Figures 2c). This graph incorporates several frequencies, seen in Supplementary Figure 2d, all being linear combinations of AB and CD frequencies, similarly to Fig. 3b in the main text. The AB frequency in this configuration is the same as the one in Supplementary Figures 2a and the additional CD frequency is $\Delta V_{MG}^{(CD)} = 1.7 \, mV$

The underlying frequencies $\Delta V_{MG}^{(AB)}$ and $\Delta V_{MG}^{(CD)}$ clearly differ by their dependence and independence (respectively) on the magnetic field (see Eq. 1,2 in the main text). In Supplementary Figure 3 we show their dependence on magnetic field, measured while the isolation-gate was un-biased (Supplementary Figure 3a) and biased (Supplementary Figure 3b). The $1/\Delta V_{MG}^{(AB)}$ (green, diamond) frequency is similar in both graphs, showing a clear linear dependence on the magnetic field, in agreement with Eq. 1. The $1/\Delta V_{MG}^{(CD)}$ frequency (blue, triangles) is recognized by its independence on magnetic field according to Eq. 2. Then, the third frequency (red, triangle), coincides with the sum of the two elementary ones $1/\Delta V_{MG}^{(CD)} + 1/\Delta V_{MG}^{(AB)}$ (cyan, circles).

**Supplementary Note 2. Capacitive model of the FPI**

**(a) A capacitive model for the Fabry-Perot Interferometer**

The FPI setup is depicted in Supplementary Figures 5 as an electrical circuit. By means of two QPCs, a quantum Hall (QH) strip is separated into (weakly coupled) three parts: left and right leads, and the FPI in the middle which can be regarded as a quantum dot (QD). As in the main text, we consider the case of filling factor $1 < \nu_B < 2$. The FPI consists of an outer edge channel (with a chiral one-dimensional model), and an interior compressible puddle in the bulk[2], separated by an incompressible $\nu = 1$ electron gas.

We consider a capacitive model, described by an equivalent electrostatic circuit, depicted in Fig. S2, which is widely accepted to describe our setup[1,3,4]. It consists of three effective capacitors, $C_{edge}$, $C_{bulk}$, $C_{eb}$, describing the electrostatic energy due to charging at the edge channel, at the bulk, and due to edge-bulk interaction, respectively (Fig. S2). The charge distribution at the edge and the bulk is dictated, in principle, by the varied perpendicular magnetic field $\delta B$, and the modulation gate $\delta V_{MG}$, and is subject to minimization of the electrostatic energy.

We first consider an initial tuning of the FPI. The initial value of the magnetic field is $B_0$. The system's ground state is described by the charge $Q_{edge} = eN_{edge}$ with $N_{edge}$ electrons occupying the LLL, and the charge $Q_{bulk} = eN_{bulk}$ with $N_{bulk}$ electrons located at the compressible bulk puddle. The edge of the FPI has the initial boundary of the incompressible $\nu = 1$ area $\bar{A}$, dictated by the chemical potential of the leads. We will argue below that this boundary may be modified by varying the magnetic field $B$ and the modulation gate voltage $V_{MG}$. In order to determine the ground state configuration $(N_{edge}, N_{bulk})$ for the $(B, V_{MG})$ plane, we first calculate the total electrostatic energy.

We note that even if the number of electron occupying the $\nu = 1$ level in the FPI remains unchanged (i.e. no transfer of electrons to/from the bulk puddle or the leads), charging of the edge can be induced by varying the magnetic field $\delta B = B - B_0$. The excess charge on the edge is then:

$$\delta q_3 - \delta q_1 = \delta Q_{edge} - e \cdot \frac{A \delta B}{\phi_0} = e \delta N_{edge} - e \cdot \frac{A \delta B}{\phi_0}. \quad (S1)$$

The compressible puddle at the center of the FPI, i.e., the "bulk", serves as an effective reservoir: it may take out or give away charge from/to the edge, then minimizing the electrostatic energy. The excess charge in the bulk is expressed as:

$$-\delta q_2 - \delta q_3 = \delta Q_{bulk} + e \cdot \frac{A\delta B}{\phi_0} = e\delta N_{bulk} + e \cdot \frac{A\delta B}{\phi_0}. \qquad (S2)$$

For a fixed $\delta N_{bulk}$, the variation $\delta B > 0$ leads to charge accumulation in the bulk of the $\nu = 1$ Landau level by $e \cdot \frac{A\delta B}{\phi_0}$. As a matter of fact, the number of electrons in the bulk of the FPI is the sum of two contributions: the electrons of which the incompressible $\nu = 1$ liquid is comprised of, and the electrons forming the compressible puddle in the middle. To simplify the notation, "bulk" here refers only to that latter contribution. The total electrostatic energy of the electric circuit is expressed as:

$$E_{total} = \frac{q_1^2}{2C_{edge}} + \frac{q_2^2}{2C_{bulk}} + \frac{q_3^2}{2C_{eb}} - q_1 \delta V_{MG} - q_2 \delta V_{MG}$$

$$= \frac{K_E}{2} e^2 \cdot \left(\delta N_{edge} - \frac{A\delta B}{\phi_0} - \frac{\alpha B_0 \delta V_{MG}}{\phi_0}\right)^2 + \frac{K_{EB}}{2} e^2 \cdot \left(\delta N_{edge} + \delta N_{bulk} - \frac{\gamma \delta V_{MG}}{|e|}\right)^2$$

$$+ \frac{K_B}{2} e^2 \cdot \left(\delta N_{bulk} + \frac{A\delta B}{\phi_0} + \frac{\alpha B_0 \delta V_{MG}}{\phi_0} - \frac{\gamma \delta V_{MG}}{|e|}\right)^2$$

$$- \frac{C_{edge} + C_{bulk}}{2} (\delta V_{MG})^2. \qquad (S3)$$

Here, $K_{EB} + K_E = \frac{C_{bulk} + C_{eb}}{D}$, $K_{EB} + K_B = \frac{C_{edge} + C_{eb}}{D}$, $K_{EB} = \frac{C_{eb}}{D}$, and $D = (C_{eb} + C_{edge})(C_{eb} + C_{bulk}) - C_{eb}^2$. In terms of the AB phase Eq. S3 reads:

$$E_{total} = \frac{K_E}{2} e^2 \cdot \left(\delta N_{edge} - \frac{\delta \phi_{AB}}{\phi_0}\right)^2 + \frac{K_{EB}}{2} e^2 \cdot \left(\delta N_{edge} + \delta N_{bulk} - \frac{\gamma \delta V_{MG}}{|e|}\right)^2 + \frac{K_B}{2} e^2$$

$$\cdot \left(\delta N_{bulk} - \left(\frac{\gamma \delta V_{MG}}{|e|} - \frac{\delta \phi_{AB}}{\phi_0}\right)\right)^2 - \frac{C_{edge} + C_{bulk}}{2} (\delta V_{MG})^2. \qquad (S4)$$

We will disregard the last two terms since they do not depend on the number of the electrons, and hence play no role in determining the ground state; evidently, dropping them results in Eq. 4 in the main text. The parameter $\alpha = \frac{C_{edge}}{B_0 |e|} \phi_0$ effectively accounts for changes of the area by MG voltage variations $\delta V_{MG}$, while keeping $\delta Q_{edge}$ unchanged, and the parameter $\gamma$ is defined as $C_{edge} + C_{bulk}$.

**(b) Relation to previous works**

In the work by Halperin *et al*[1] the energy of the system was formulated as follows:

$$E = \frac{K_I}{2} (\delta n_I)^2 + \frac{K_L}{2} (\delta n_L)^2 + K_{IL} \delta n_I \delta n_I. \qquad (S5)$$

The first two terms in this expression are physically (up to the coefficients) equivalent to the first and third terms in Eq. S3 & S4; namely:

$$\delta n_I = \delta Q_{edge} - e \cdot \frac{\delta \phi_{AB}}{\phi_0} = e \cdot \frac{B \delta A_{res}}{\phi_0}. \qquad (S6.a)$$

$$\delta n_L = \delta Q_{bulk} + e \cdot \frac{\delta \phi_{AB}}{\phi_0} - \gamma \delta V_{MG}. \qquad (S6.b)$$

$$\delta Q_{tot} = \delta n_I + \delta n_L + \gamma \delta V_{MG}. \qquad (S6.c)$$

Now physically the interpretation of the third term in the energy above differs from that of the first term in our energy; $K_{IL} \cdot \delta n_I \delta n_L$ represents an effective interaction between the edge and the bulk, while $\frac{K_{EB}}{2}(\delta Q_{tot})^2$ represents an effective total charging energy. The relation between the different coefficients reads:

$$K_{IL} = K_{EB}. \qquad (S7.a)$$

$$K_I = K_{EB} + K_E. \qquad (S7.b)$$

$$K_L = K_{EB} + K_B. \qquad (S7.c)$$

And we can identify at ease that $\Delta = \xi$, where $\Delta \in [0,1]$ is the parameter that denotes the regime in Halperin et al.'s model[1].

### (c) Calculation of the conductance

In this section, we derive an expression of the conductance through a FPI. For a fixed $\delta N_{bulk}$, the total transmission through the FPI is expressed as $T_{\delta N_{bulk}} = |\tau_{\delta N_{bulk}}|^2$ with the total transmission amplitude $\tau_{\delta N_{bulk}} = t_l t_r (1 + \sum_{n=1}^{\infty} (r_l r_r)^n e^{2\pi \cdot i \cdot n \delta \phi_{tot}/\phi_0})$, considering the summation over all possible number of roundtrips $n$ of an interfering electron. Here $r_l$ ($r_r$) and $t_l$ ($t_r$) are the reflection and transmission amplitudes of the left (right) QPC, respectively. This total transmission is simplified as

$$T_{\delta N_{bulk}} = \frac{T_L T_R}{1 + R_L R_R - 2\sqrt{R_L R_R} \cos(2\pi \delta \phi_{tot}/\phi_0)}, \qquad (S8)$$

where $T_{L,R} = 1 - R_{L,R} = |t_{l,r}|^2 = 1 - |r_{l,r}|^2$. In the single particle picture, $\delta \phi_{tot}$ is determined by the ratio of the highest energy level ($\epsilon = -\frac{\delta \phi_{AB}}{\phi_0} \Delta + K_{EB} \left( e \cdot \delta N_{bulk} + e \cdot \frac{\delta \phi_{AB}}{\phi_0} - \gamma \delta V_{MG} \right)$) below Fermi energy to the level spacing ($\Delta = K_E + K_{EB}$),

$$\frac{\delta \phi_{tot}}{\phi_0} \equiv -\frac{\epsilon}{\Delta} = \frac{\delta \phi_{AB}}{\phi_0} - \xi \left( \delta N_{bulk} + \frac{\delta \phi_{AB}}{\phi_0} - \frac{\gamma \delta V_{MG}}{|e|} \right). \qquad (S9)$$

Two comments are due here: Eq. S9 coincides with a single-particle analysis of transmission through the FPI. There are two facets of many-body physics which are neglected here. (i) Putting aside higher order processes in the tunneling amplitudes $\{\Gamma\}$, we can neglect inelastic transmission processes that leave a trace on the FPI (By contrast, cf. Ref. 5, there are scenarios where inelastic contributions are of the same order as elastic ones). (ii) Assuming $K_{EB} \ll K_B$, we may ignore the back action of varying the number of electrons in the bulk due to a change of the edge configuration.

When $\delta N_{bulk}$ fluctuates, Eq. S8 should be modified as $\langle T \rangle = \sum_{\delta N_{bulk}} P_{\delta N_{bulk}} T_{\delta N_{bulk}}$, where $P_{\delta N_{bulk}}$ is the probability that the number of the electrons in the bulk is $\delta N_{bulk}$. In order to calculate $P_{\delta N_{bulk}}$, we first start with the occupation number $\langle \delta N_{bulk} \rangle$. Within the additional approximation of $\Gamma_{bulk} \ll (K_B + K_{EB})$, while keeping $\Gamma_{bulk} > 0$, the bulk system can be treated as a two-state system, an occupied ($o$) and unoccupied ($u$) states; this terminology refers to the closest level to the Fermi level; fluctuation of $\delta N_{bulk}$ that go further than $\pm 1$ are neglected. The bulk state is then expressed as $\sqrt{P_u}|u\rangle + \sqrt{P_o}|o\rangle$, where $P_o$ ($P_u$) is the probability for an (un)occupied state of the said level with $P_o = \langle \delta N_{bulk} \rangle - F[\langle \delta N_{bulk} \rangle]$ ($P_u = 1 - P_o$) and $F[x]$ is the integer part of $x$. The occupation number $\langle \delta N_{bulk} \rangle$ is determined by the integral of the single particle bulk states centered at $\epsilon_n = (K_{bulk} + K_{EB})(n + \frac{\delta \phi_{AB}}{\phi_0} - \frac{\gamma \delta V_{MG}}{|e|})$ over the energy as:

$$\langle \delta N_{bulk} \rangle \equiv \sum_{n=0}^{\infty} \frac{1}{\pi} \int_{-\infty}^{0} \frac{\Gamma_{bulk}}{(\epsilon - \epsilon_n)^2 + \Gamma_{bulk}^2} d\epsilon$$

$$= \sum_{n=0}^{\infty} \left( \frac{1}{2} - \tan^{-1} \left[ \frac{K_B + K_{EB}}{\Gamma_{bulk}} \left( n + \frac{\delta \phi_{AB}}{\phi_0} - \frac{\gamma \delta V_{MG}}{|e|} \right) \right] \right). \quad (S10)$$

This conductance is related with $\langle T \rangle$, which is the average over the different possible values of $\delta N_{bulk}$ with the corresponding probabilities:

$$G = \frac{e^2}{h} \langle T \rangle = \frac{e^2}{h} \left( P_u T_{F[\langle \delta N_{bulk} \rangle]} + P_o T_{F[\langle \delta N_{bulk} \rangle]+1} \right). \quad (S11)$$

This equation is used for drawing the Figure 4.b and Supplemental Figures 8.c and 9.b.

**(d) Charge stability diagram**

In order to formally express the charge stability diagram we first optimize the energy in Eq. S4 with respect to both $\delta A_{res}$ and $\delta Q_{in}$ at the same, resulting in:

$$B \delta A_{res} = 0. \quad (S12.a)$$

$$\delta Q_{bulk} = -e \cdot \frac{A\delta B}{\phi_0} - \left(e \cdot \frac{\alpha B}{\phi_0} - \gamma\right)\delta V_{MG}. \quad (S12.b)$$

Every time these two equations are satisfied, all regimes coincide since both the dipole and the charging energies are at local minima (in fact, it is clear that plugging Eq. S12 in Eq. S4 results in $\delta E = 0$). The charge-states which satisfy Eq. 10 $(\delta N_{edge}, \delta N_{bulk}) = (n, m)$ produce the vectors:

$$(\delta B, \delta V_{MG}) = \left(\Delta B^{(AB)}\left(n - \frac{\Delta V_{MG}^{(CD)}}{\Delta V_{MG}^{(AB)}} \cdot (n+m)\right), \quad \Delta V_{MG}^{(CD)} \cdot (n+m)\right), \quad (S13)$$

which describe the vectors of the charge-stability diagram. Similar to a Bravais lattice, these vectors may be spanned by two:

$$\boldsymbol{a}_1 = \left(0, \Delta B^{(AB)}\right), \quad (S14.a)$$

$$\boldsymbol{a}_2 = \left(\Delta V_{MG}^{(CD)}, -\Delta B^{(AB)} \frac{\Delta V_{MG}^{(CD)}}{\Delta V_{MG}^{(AB)}}\right), \quad (S14.b)$$

so that the lattice is given by $\boldsymbol{R}_{n,m} = \sum_{n,m} n \cdot \boldsymbol{a}_1 + m \cdot \boldsymbol{a}_2$. Hence the reciprocal lattice is given by the four equations $\boldsymbol{a}_j \cdot \boldsymbol{b}_i = 2\pi\delta_{ij}$ which results in the:

$$\boldsymbol{b}_1 = 2\pi \cdot \boldsymbol{\omega}_{AB} = \left(\frac{2\pi}{\Delta V_{MG}^{(CD)}}, \frac{2\pi}{\Delta B^{(AB)}}\right), \quad (S15.a)$$

$$\boldsymbol{b}_2 = 2\pi \cdot \boldsymbol{\omega}_{CD} = \left(\frac{2\pi}{\Delta V_{MG}^{(CD)}}, 0\right), \quad (S15.b)$$

which are indeed the underlying AB and CD frequencies, as anticipated.

Now it is clear from Eq. S14 above that, the vectors spanning the charge stability diagram do not depend on the regime. Nevertheless, the different regimes' charge stability diagram (CSD) differ by the shape of the unit-cells, as we show in the different asymptotic cases in the following section. The full CSD including the structures of the unit-cells is obtained by finding $(\delta N_{edge}, \delta N_{bulk})$ which minimize the energy (Eq. S4). In this way, Supplementary Figures 7 & 8a, discussed in what follows, are simulated.

**(d) The Aharonov-Bohm and Coulomb-dominated regime**

First we discuss the two distinct regimes of behavior of the FPI: the Aharonov-Bohm (AB) regime and the Coulomb-dominated (CD) regime[1,4]. These are distinguished by the slope of their equi-phase lines in the 2D conductance as function of magnetic field $\delta B$ and modulation gate voltage $\delta V_{MG}$ (the so-called pajama patterns). In the AB regime, the equi-phase lines have a negative slope; when the magnetic field increases, the phase remains invariant as the modulation gate voltage (and thus the area) decreases. On the other hand, in the CD regime, equi-phase lines follow a non-negative slope. The different regimes can be retrieved by examining the asymptotic limits of the energy given in Eq. S3; we shall discuss, for each of the asymptotic limits, its physical interpretation and its charge stability diagram.

*AB regime*: Two asymptotic choices of parameters yield this regime, discussed separately below. In both cases $K_E \gg K_{EB}$, or, namely, $\xi = 0$.

(i) A truly non-interacting system, expressed in terms of the effective capacitances: $C_{edge} \to \infty, C_{bulk} \to \infty, C_{eb} \to \infty$ (see Supplementary Figure 7a). In this regime, $K_B = 0, K_{EB} = 0$, and $K_E$ corresponds to the level spacing of the single-particle energies, a scale which is not included in Eq. S3. $K_E$ is then determined by the slope of the confining potential at the Fermi level, as well as the magnetic field $B$, and the area $\bar{A}$. Since $K_B = 0$, the charge imbalance in the bulk does not play a role in determining the ground state configuration so that the CSD describes the charge state of the edge solely.

(ii) More interestingly, and experimentally relevant, it is possible to obtain a negative slope of the constant phase lines for an interacting system (see Supplementary Figure 7b) by requiring $C_{bulk} \gg C_{eb}$ and $K_E \gg K_{EB}$. This result is equivalent to our measurements with the AB-dominated device shown in Fig. 2a.

*CD regime*: Here too, we consider two asymptotic scenarios.

(iii) $C_{bulk} \ll C_{eb}$, and $C_{edge} > C_{bulk}$ ($K_E \ll K_{ebEB}$, $K_E < K_B$) (see Supplementary Figure 7c). In this regime, the lines of constant phase and the lines of conductance maxima are vertical in the $(\delta B, \delta V_{MG})$ plane. If one crosses the blue lines seen in Fig. S3c, redistribution of the electrons between the bulk and the edge takes place. This result is equivalent to our measurements with the CD device shown in Fig. 2b.

(iv) $C_{bulk}, C_{edge} \ll C_{eb}$ ($K_E, K_B \ll K_{EB}$) (see Supplementary Figure 7d). While the lines of constant phase and the lines of conductance maxima are vertical in the scenario (i) of the CD regime (cf. Supplementary Figure 7c), here vertical boundaries consist of alternating black lines (where $\delta N_{bulk}$ is changed by $\pm 1$ with constant $\delta N_{edge}$) and the red lines (where $\delta N_{edge}$ is changed by $\pm 1$ with constant $\delta N_{bulk}$).

### (e) The intermediate regime

Finally, we discuss the intermediate regime[6] between the asymptotic AB and CD regimes. As shown in Supplementary Figure 8a, the conductance peak (red) lines with a negative slope (the signature of the AB regime) separate states with different total number of electrons, $\delta N_{bulk} + \delta N_{edge}$ (the signature of the CD regime). The intermediate regime can be described by the parameters $0 < \frac{K_{EB}}{K_E+K_{EB}} < 1$, as compared with $\frac{K_{EB}}{K_E} = 0$ for the AB regime and $\frac{K_{EB}}{K_E+K_{EB}} = 1$ for the CD regime. The conductance in the $(\delta B, \delta V_{MG})$ plane based on the capacitive model is plotted in Supplementary Figure 8c (low transmission) and Supplementary Figure 9b (high transmission), and they are compared to the experimental results (Supplementary Figure 8b and Supplementary Figure 9a, respectively).

### Supplementary Note 3. High transmission approximation

#### (a) Conductance

In the high transmission limit (namely, $T \lesssim 1$), and assuming $\Gamma_{bulk} = 0$, we may approximate:

$$G(T \lesssim 1) = \frac{e^2}{h} T^2 (1 + R^2 \cos(2\pi \cdot \delta\phi_{tot}/\phi_0)). \qquad (S16)$$

This expression is used for the theoretical graphs in Fig. 5.

Most generally, for any setting of our two external knobs $\delta B$ & $\delta V_{MG}$, the system optimizes its energy by setting $\delta A_{res}$ & $\delta Q_{bulk}$ (or equivalently $\delta Q_{edge}$ & $\delta Q_{bulk}$), according to the energy given in Eq. S4. The high transmission limit is translated into relaxing the assumption of quantization of $\delta Q_{edge}$, which is made in Supplementary Note 2. In this way we can retrieve the total phase in a straightforward fashion by optimizing the energy with respect to $\delta A_{res}$, for a given value of $\delta Q_{bulk} \in [0, \pm 1, \pm 2, \ldots]$:

$$\frac{\partial \delta E}{\partial \delta A_{res}} = K_C \cdot (\delta Q_{tot} - \gamma \delta V_{MG}) \cdot \frac{B}{\phi_0} + K_E \cdot \left(\delta Q_{edge} - e \cdot \frac{\delta\phi_{AB}}{\phi_0}\right) \cdot \frac{B}{\phi_0}, \qquad (S17)$$

setting now $\frac{\partial \delta E}{\partial \delta A_{res}} = 0$ we get:

$$e \cdot \frac{B \delta A_{res}}{\phi_0} = \xi \left(\gamma \delta V_{MG} - e \cdot \frac{\delta\phi_{AB}}{\phi_0}\right) - \xi \cdot \delta Q_{bulk}. \qquad (S18)$$

Now, once plugged into the total phase we get:

$$\frac{\delta\phi_{tot}}{\phi_0} = \frac{\delta\phi_{AB}}{\phi_0} + \xi\left(\frac{\gamma\delta V_{MG}}{e} - \frac{\delta\phi_{AB}}{\phi_0}\right) - \xi \cdot \frac{\delta Q_{bulk}}{e}, \quad (S19a)$$

or:

$$\frac{\delta\phi_{tot}}{\phi_0} = (1-\xi) \cdot \frac{\delta\phi_{AB}}{\phi_0} + \xi \cdot \frac{\gamma\delta V_{MG}}{e} - \xi \cdot \frac{\delta Q_{bulk}}{e}. \quad (S19b)$$

Now the first two terms are continuously-varying variables which describe the interference phase of electrons in the outer EC for a constant bulk-charge $\delta Q_{bulk}$; while the third term describes the phase-jump which occurs once $\delta Q_{bulk}$ changes by $\pm e$ (as further explained below).

We stress here that the transmission $T$ does not affect the value of $\xi$, but only the shape of the unit-cells in the CSD as well as the clarity of its lattice-like, via SP states broadening, as shown in Supplementary Figure 5 with both experimental data and theoretical.

**(b) Phase-jump lines**

As seen in Supplementary Figure 7a, the lines along which $\delta Q_{bulk}$ increments by $\pm e$ are a set of broken lines ('zig-zags'). These zig-zags consist of two types of segments; the first (dark-blue) describe a processes of charging the bulk $(\delta Q_{bulk}, \delta Q_{edge}) \rightarrow (\delta Q_{bulk} \pm e, \delta Q_{edge})$; and the second (light-blue) describe 'reshuffling' of the charge between the edge and the bulk $(\delta Q_{bulk}, \delta Q_{edge}) \rightarrow (\delta Q_{bulk} \pm e, \delta Q_{edge} \mp e)$.

Nonetheless, for higher transmission probabilities, these zig-zags become straight lines. These result can be easily obtained be rewriting the energy in the following form:

$$E_{total} = \frac{K_E}{2}\left(\delta Q_{edge} - e \cdot \frac{\delta\phi_{AB}}{\phi_0} + \frac{K_{EB}}{K_E} \cdot \left(\delta Q_{bulk} - \left(\gamma\delta V_{MG} - e \cdot \frac{\delta\phi_{AB}}{\phi_0}\right)\right)\right)^2$$

$$+ \frac{\left(K_E + \frac{K_{EB}^2}{K_E}\right)}{2}\left(\delta Q_{bulk} - \left(\gamma\delta V_{MG} - e \cdot \frac{\delta\phi_{AB}}{\phi_0}\right)\right)^2 \quad (S20)$$

Since at the high transmission limit we consider $\delta Q_{edge}$ to be a continuous variable, it is evident that the first term in this energy is zero at all times (since $\delta Q_{edge}$ appears only within it. Thus the energy effectively depends on the second term only, which is clearly minimized by setting $\delta Q_{bulk} = \left(\gamma\delta V_{MG} - e \cdot \frac{\delta\phi_{AB}}{\phi_0}\right)$. Writing this expression for any $\delta Q_{bulk} = ne$ with $n$ an integer, results in the line in the $B - V_{MG}$ plane along which having exactly $\delta Q_{bulk} = ne$ is optimal. Thus, very intuitively, the lines along which $\delta Q_{bulk}$ changes by $\pm 1$ may be obtained by setting $\delta Q_{bulk} = (n + \frac{1}{2}) \cdot e$:

$$\frac{A}{\phi_0} \delta B + \left(\frac{\alpha B}{\phi_0} - \frac{\gamma}{e}\right) \delta V_{MG} = n + \frac{1}{2}, \quad (S21a)$$

or:

$$\frac{\gamma \delta V_{MG}}{e} - \frac{\delta \phi_{AB}}{\phi_0} = n + \frac{1}{2}, \quad (S21b)$$

By taking a variation on this equation resulting in $\Delta n = 1$ we obtain the PJ lines 2D frequencies (assuming high QPCs transmission):

$$1/\Delta B_{PJ} = A/\phi_0, \quad (S22a)$$

$$\frac{1}{\Delta V_{PJ}} = \frac{\alpha B}{\phi_0} - \frac{\gamma}{e}, \quad (S22b)$$

or, in terms of the underlying AB and CD frequencies:

$$1/\Delta B^{(PJ)} = 1/\Delta B^{(AB)}, \quad (S23a)$$

$$1/\Delta V_{MG}^{(PJ)} = 1/\Delta V_{MG}^{(AB)} - 1/\Delta V_{MG}^{(CD)}, \quad (S23b)$$

**Supplementary Figures**

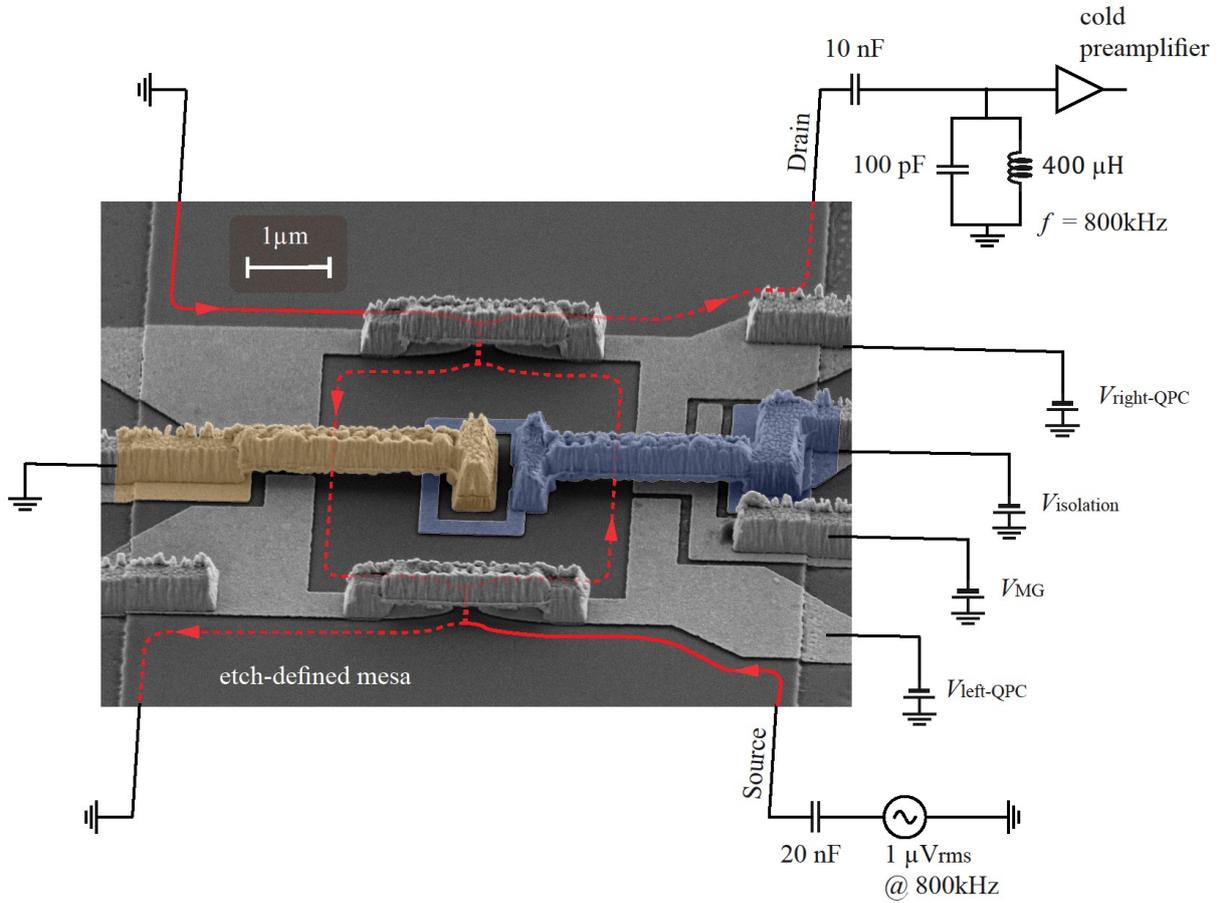

**Supplementary Figure 1: SEM images and illustration of a FPI with a center ohmic contact surrounded by an isolation-gate.** The regime of this device can be tuned by varying the voltage on the isolation gate $V_{isolation}$; it can be tune to either the pure AB regime, by setting $V_{isolation} = 0$, or to an intermediate regime by setting a low enough voltage resulting in depletion of the electron-gas below it. We note that a pure CD regime cannot be reached with this device.

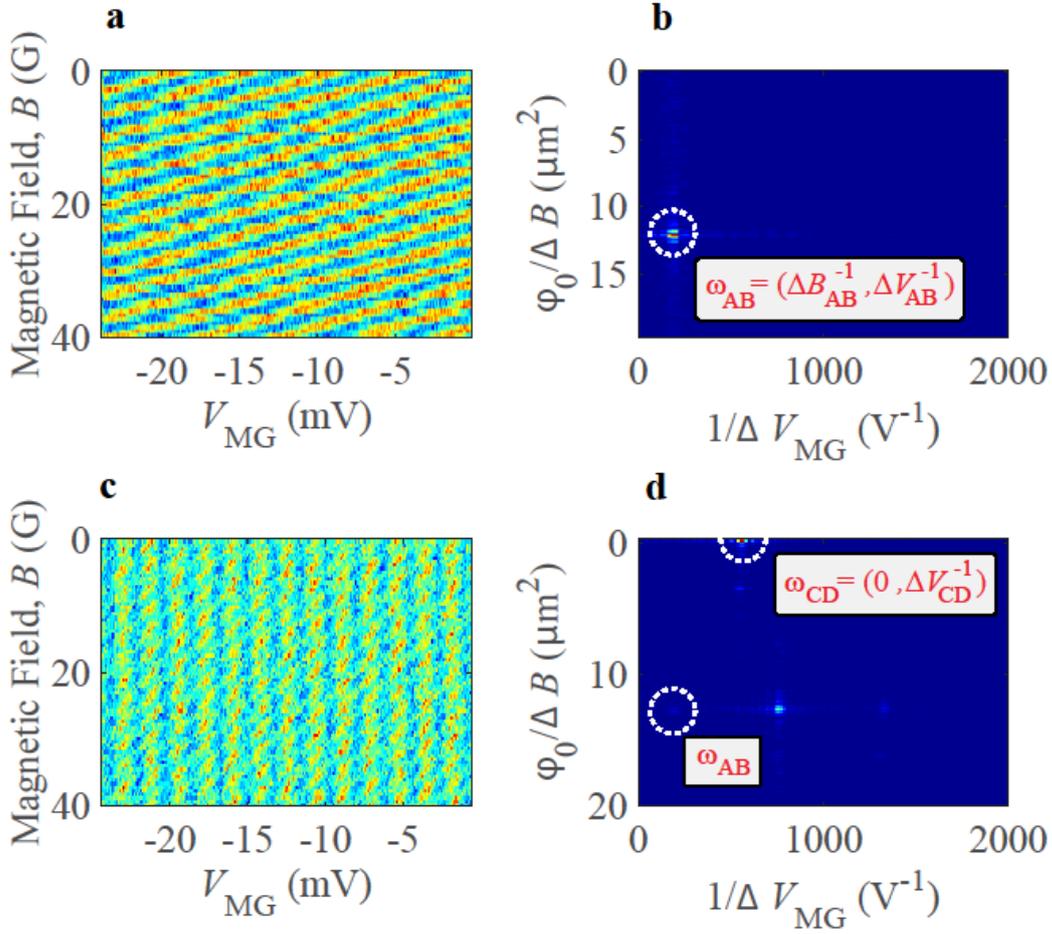

**Supplementary Figure 2: Conductance measurements as function of the modulation-gate voltage $V_{MG}$ and magnetic field $B$, measured with the 12 $\mu m^2$ device with an isolation-gate around its center ohmic-contact shown in Fig. 1d. a**, 2D conductance measured while the isolation-gate is unbiased. This plot shows clear AB behavior. **b**, FFT for a; A single frequency $\omega_{AB}$ is observed, in agreement with Eq. 1. **c**, 2D conductance measured while the isolation-gate is biased so that the 2DEG beneath it is depleted. **d**, FFT for c; several frequencies are observed, all being linear combinations of the AB and CD frequencies, $\omega_{AB}$ and $\omega_{CD}$ respectively. We note that $\omega_{AB}$ in the two plots takes the same value, as anticipated. Dependence of these frequencies on magnetic field is detailed in Supplementary Figure 3.

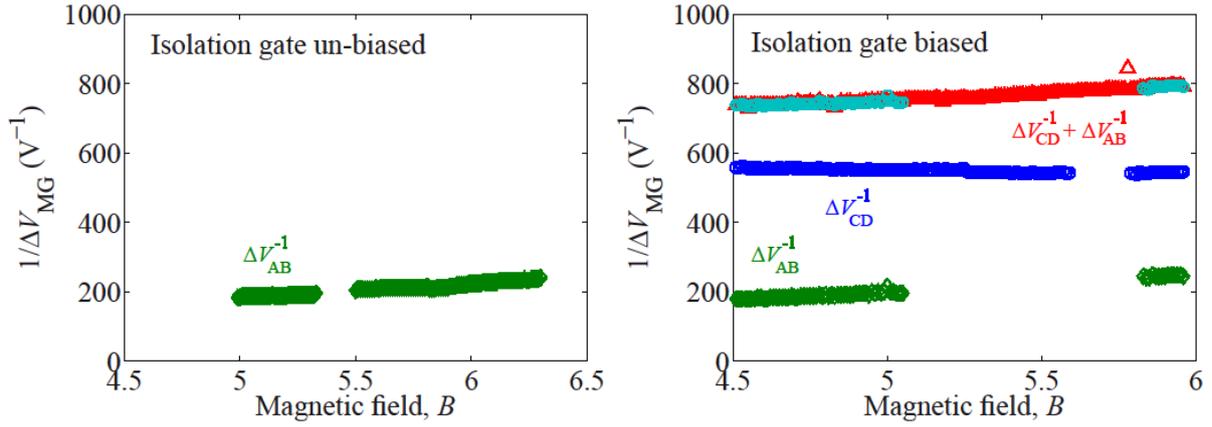

**Supplementary Figure 3: Aharonov-Bohm and Coulomb-dominated frequencies ($1/\Delta V_{MG}^{(AB)}$ and $1/\Delta V_{MG}^{(CD)}$) as function magnetic field, measured with the 12 $\mu m^2$ device with an isolation-gate around its center ohmic contact shown in Fig. 1d. a**, AB frequency measured while the isolation gate is unbiased. **b**, AB and CD frequencies measured while the isolation-gate is pinched-off, giving rise to the intermediate regime. The AB frequency (green, diamond) is easily distinguished according to the graph in a. The CD frequency is distinguished by its independence on magnetic-field according to Eq. 2. The third measured frequency (red, triangles) coincides with the some of the two (cyan, circles).

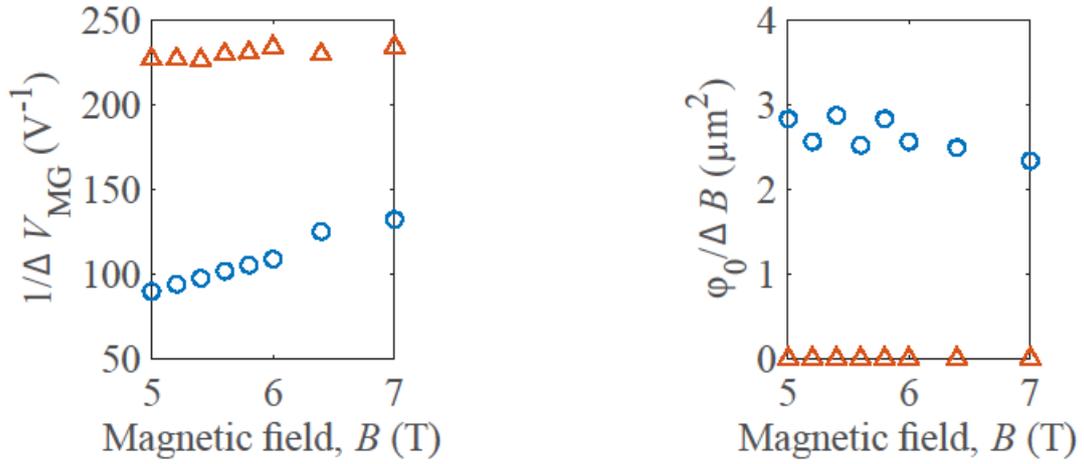

**Supplementary Figure 4**: **Aharonov-Bohm and Coulomb-dominated frequencies ($1/\Delta V_{MG}^{(AB)}$ and $1/\Delta V_{MG}^{(CD)}$) as function magnetic field, measured with the $2.6\ \mu m^2$ device with an center ohmic contact in its center, shown in Fig. 1c. a**, Modulation-gate frequencies with respect to magnetic field; AB frequency $1/\Delta V_{MG}^{(AB)}$ (blue, circles) and CD frequency $1/\Delta V_{MG}^{(CD)}$ (red, triangles). While the first depends linearly on magnetic field, the second doesn't, as expected according to Eq. 1 & 2. **b**, Magnetic field frequencies (corresponding to periods in the Gauss scale) with respect to magnetic field (in the Tesla scale); AB frequency $1/\Delta B_{MG}^{(AB)}$ (blue, circles) and CD frequency $1/\Delta V_{MG}^{(CD)}$ (red, triangles). While the first has a finite value representing the device's area, the second doesn't; both of the frequencies do not show a dependence linearly on magnetic field, as expected according to Eq. 1 & 2.

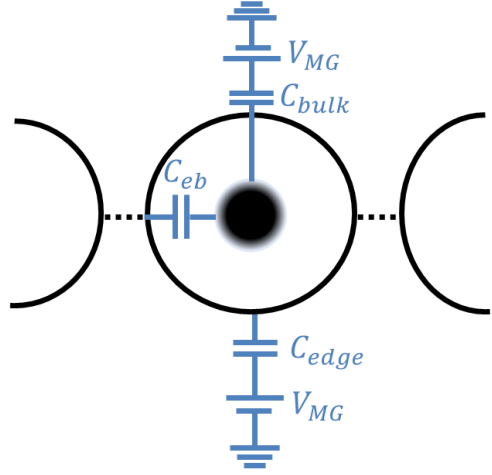

**Supplementary Figure 5: A Fabry-Perot interferometer at filling factor $1 < \nu_B < 2$.** The system consists of an edge of the $\nu_B = 1$ Landau level and a compressible puddle of charge in the bulk, separated by an incompressible $\nu = 1$ region. It is effectively modeled by three capacitors, $C_{edge}, C_{bulk}, C_{eb}$, describing the electrostatic energy due to charging at the edge, at the bulk, and due to edge-bulk interaction.

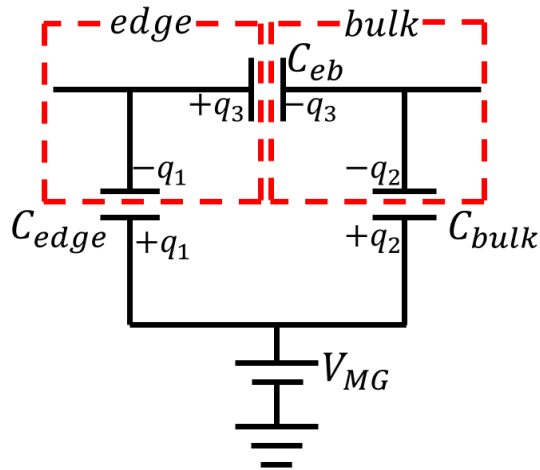

**Supplementary Figure 6: An electric circuit model describing a quantum Hall FPI (or equivalently QD) with filling factor $1 < \nu_B < 2$.** The system consists of an edge of the lowest Landau level (hereafter "edge") which undergoes the interfering trajectories, and the bulk (denoted as "bulk"). The edge of the interfering lowest Landau level is capacitive coupled to the bulk through $C_{eb}$, and to the MG (located outside the interfering edge) through $C_{edge}$. The bulk is coupled to the MG as well, through $C_{bulk}$. The excess charge on the edge is expressed as $\delta q_3 - \delta q_1$, while the excess charge in the bulk is expressed as $-\delta q_3 - \delta q_2$.

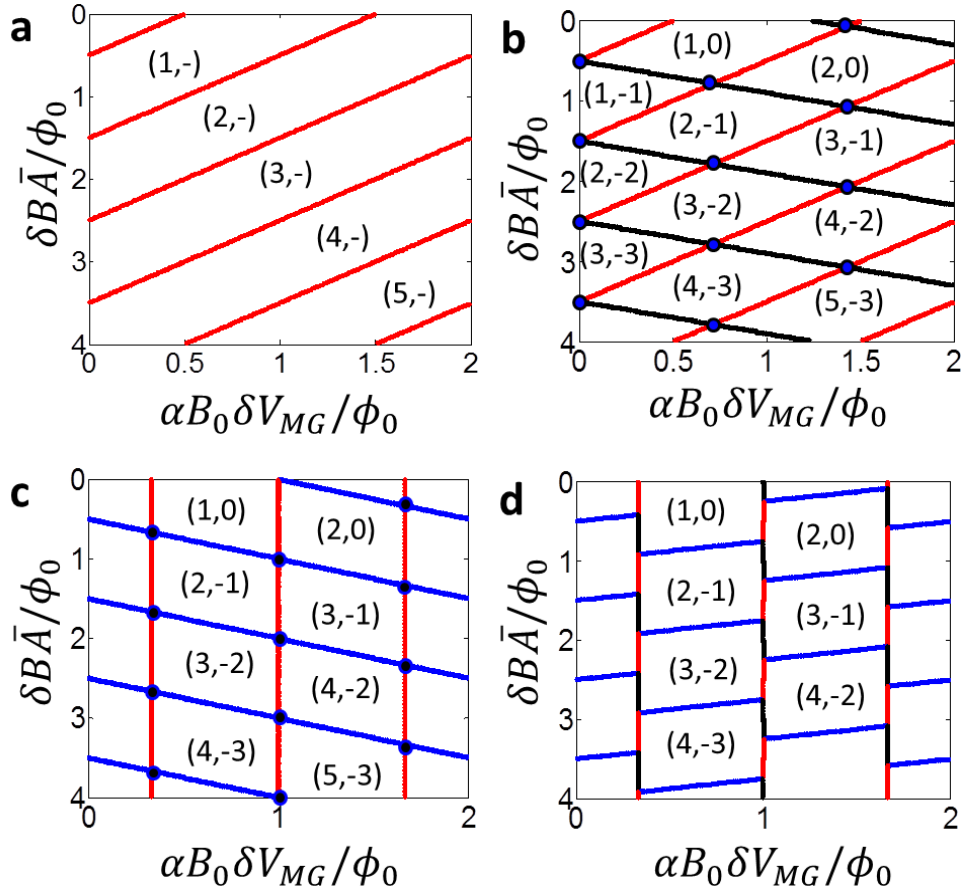

**Supplementary Figure 7: Charge stability diagrams of the FPI in the $(\delta B, \delta V_{MG})$ plane. a, b, c, d,** Charge stability diagrams for scenario (i), (ii), (iii) and (iv) respectively. The numbers in parentheses indicate the charge state $(\delta N_{edge}, \delta N_{bulk})$. Note that in **a**, the number of electrons in the bulk is not well-defined (no requirement for charge neutrality). Here *red lines* represent degeneracy between $(\delta N_{edge}, \delta N_{bulk})$ and $(\delta N_{edge} + 1, \delta N_{bulk})$, corresponding to conductance peaks. *Blue lines* represent edge-bulk charge reshuffling (inducing a phase jumps), i.e., $(\delta N_{edge}, \delta N_{bulk}) \to (\delta N_{edge} + 1, \delta N_{bulk} - 1)$. *Black lines* represent degeneracy between $(\delta N_{edge}, \delta N_{bulk})$ and $(\delta N_{edge}, \delta N_{bulk} + 1)$. In **b**, $\frac{K_{bulk}}{K_{edge}} = 2.5$ and $\frac{K_{eb}}{K_{edge}} = 0$, in **c**, $\frac{K_{bulk}}{K_{eb}} = 1.5$ and $\frac{K_{edge}}{K_{eb}} = 0$, and in **d**, $\frac{K_{edge}}{K_{eb}} = \frac{K_{bulk}}{K_{eb}} = 0.001$. In all cases, $\frac{\gamma \phi_0}{\alpha B_0 |e|} = 1.5$. We didn't constrain the values of $K_{edge}$ and $K_{bulk}$ depending on capacitances, taking into account single particle energy level spacing in reality.

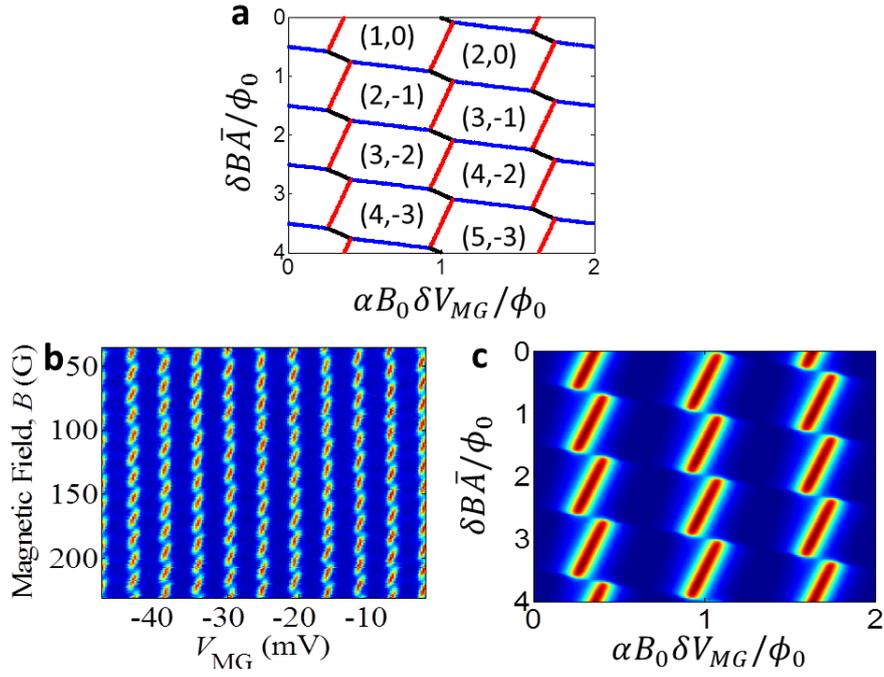

**Supplementary Figure 8: a**, Stability diagram of the QD in the intermediate regime (between the Aharonov-Bohm regimes and Coulomb-dominated regimes). Conductance is shown in the $(\delta B, \delta V_{MG})$ plane in **b**, experiment and **c**, theory based on our capacitive model. In **a** and **c**, we use parameters $\frac{K_{bulk}}{K_{edge}} = 7$, $\frac{K_{eb}}{K_{edge}} = 3$, and $\frac{\gamma \phi_0}{\alpha B_0 |e|} = 1.5$. In **c**, $\frac{\Gamma_{bulk}}{K_{edge}} = 0.0125$ and $R_L = R_R = 0.6$.

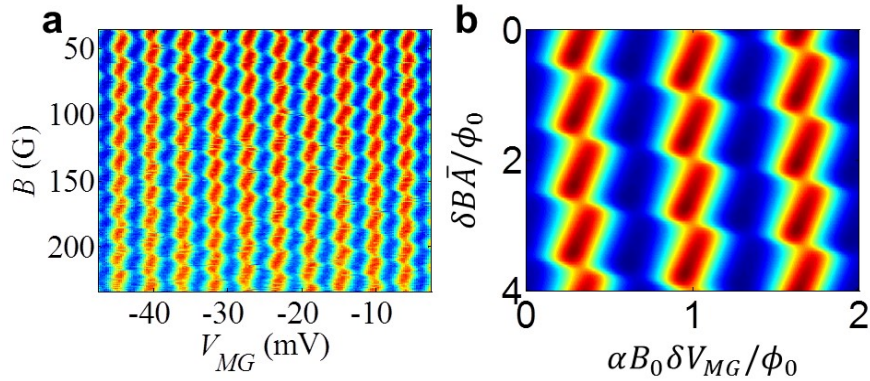

**Supplementary Figure 9:** The conductance in the $(\delta B, \delta V_{MG})$ plane in **a**, experiment and **b**, theory. In **b**, we use parameters $\frac{K_{bulk}}{K_{edge}} = 7$, $\frac{K_{eb}}{K_{edge}} = 3$, $\frac{\gamma \phi_0}{\alpha B_0 |e|} = 1.5$, $\frac{\Gamma_{bulk}}{K_{edge}} = 0.375$ and $R_L = R_R = 0.2$.

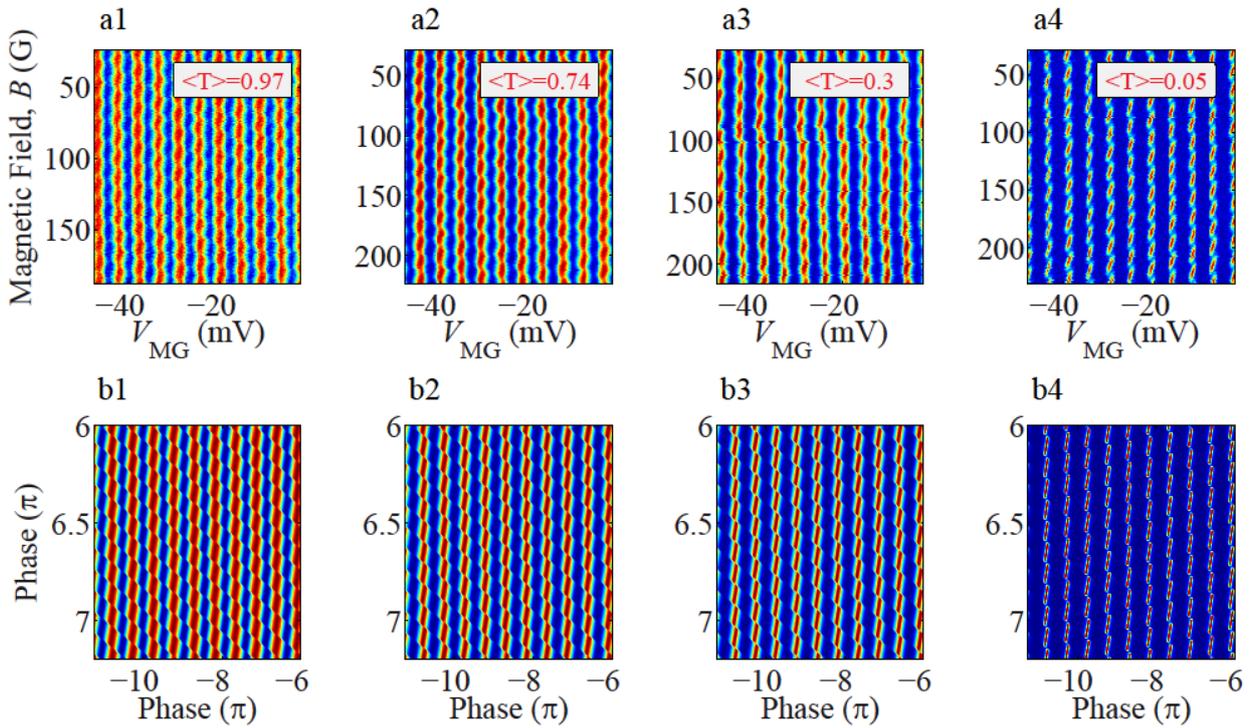

**Supplementary Figure 10**: **Effect of the transmission on the modified-AB pajama.** We shows the evolution of the pajama plot as the transmission of the FPI's QPCs are being increased. Noticeably, even when scanning the whole range of transmission from $<T> = 0.05$ to $<T> = 0.97$ the qualitative picture remains. Specifically we stress that the modified-AB frequencies do not change over this range, keeping the values given in Fig. 4a. **a1-a4**, Measured conductance as function of the modulation-gate voltage $V_{MG}$ and magnetic-field $B$ for different values of the average transmission. **b1-b4**, Calculations of the conductance oscillations (for details see Supplementary Note 3).

**Supplementary References**